\documentclass[iop,apj,onecolumn,appendixfloats]{emulateapj}

\newcommand{\bbb}{\bigskip}
\newcommand{\be}{\begin{eqnarray}}
\newcommand{\bit}{\begin{itemize}}

\def\bkt#1{\left(#1\right)} % normal brackets
\def\bkts#1{\left[#1\right]} % square brackets
\def\bkta#1{\langle#1\rangle} % angle brackets

\def\diff#1#2{{d{#1}\over d{#2}}}
\def\ee{\end{eqnarray}}
\def\eg{\textit{e.g.} }
\newcommand{\eit}{\end{itemize}}
\def\etal{\textit{et al.} }

\def\fnl{f_{\mbox{\scriptsize NL}}}
\def\ff{\phantom{.}}

\def\gnl{g_{\mbox{\scriptsize NL}}}

\def\ie{\textit{i.e.}}
\def\iee{\textit{i.e. }}
\newcommand{\ii}{\textit}

\def\lab{\label}

\newcommand{\liner}{\centerline{\vrule height 0.01in width 4.4in}}

\newcommand{\mb}{\mathbf}
\def\mc#1{\mathcal{#1}}

\def\ng{non-Gaussianity }
\newcommand{\no}{\noindent}
\newcommand{\nn}{\nonumber}

\def\pr{\prime}
\def\re#1{(\ref{#1})}

\def\sfs{\mathsf{s}}
\def\sfk{\mathsf{k}}
\newcommand{\sss}{\smallskip}
\def\sub#1{_{\mbox{\scriptsize{#1}}}}
\def\super#1{^{\mbox{\scriptsize{#1}}}}

\usepackage{amsmath}
\usepackage{amsfonts}
\usepackage{amssymb}
\usepackage{amsthm}
\usepackage{graphicx}
\usepackage{epsfig}
\usepackage{hyperref}
\begin{document}
\title{A study of high-order non-Gaussianity \\ with applications to massive clusters and large voids}
\author{Sirichai Chongchitnan}
\author{Joseph Silk}
\affil{Oxford Astrophysics \\ Denys Wilkinson Building, Keble Road, Oxford, OX1 3RH.}

\email{siri@astro.ox.ac.uk}

\begin{abstract}
The statistical meaning of the local non-Gaussianity parameters $\fnl$ and $\gnl$ is examined in detail. Their relations to the skewness and the kurtosis of the probability distribution of density fluctuations are shown to obey simple fitting formulae, accurate on galaxy-cluster scales. We argue that the knowledge of $\fnl$ and $\gnl$ is insufficient for reconstructing a well-defined distribution of density fluctuations. However, by weakening the statistical significance of $\fnl$ and $\gnl$, it is possible to reconstruct a well-defined pdf by using a truncated Edgeworth series. We give some general guidelines on the use of such a series, noting in particular that 1) the Edgeworth series cannot represent models with nonzero $\fnl$, unless $\gnl$ is nonzero also, 2) the series cannot represent models with $\gnl<0$, unless some higher-order non-Gaussianities are known. Finally, we apply the Edgeworth series to calculate the effects of $\gnl$ on the abundances of massive clusters and large voids. We show that the abundance of voids may generally be more sensitive to high-order non-Gaussianities than the cluster abundance.

\end{abstract}


\maketitle

\section{Introduction}

One of the great challenges for 21st-century cosmology is to understand the statistical distribution of primordial seeds that eventually grew to become large-scale structures. If these primordial seeds, or density fluctuations, were laid down by a simple inflationary mechanism consisting of a single scalar field, the initial distribution is expected to be very close to Gaussian (see \eg \cite{bartolo} or \cite{chen} for reviews). However, primordial non-Gaussianity can be large in more complex models such as multi-field inflation \citep{rigopoulos,byrnes}, brane inflation \citep{chen2,langlois} or in the curvaton model \citep{bartolo2,sasaki}. 

Given a physical model that generates primordial density fluctuations, it has become common practice to calculate the `local' type of non-Gaussianity parametrized by constants $\fnl$ and, less commonly, $\gnl$, defined by the expansion of the non-linear Newtonian potential
\be \Phi = \phi + \fnl(\phi^2-\bkta{\phi^2})+\gnl \phi^3+\ldots,\ee
where $\phi$ is a Gaussian random field. Observational constraints on $\fnl$  from the cosmic microwave background (CMB) anisotropies are currently consistent with $\fnl=32\pm42$ ($2\sigma$)\citep{komatsu}. Constraints on $\gnl$ are much weaker, with $|\gnl|\lesssim6\times10^5$ reported by \cite{vielva}. These limits should improve by at least an order of magnitude with results from the Planck satellite\footnote{\texttt{http://planck.cf.ac.uk}}.

But what do these numbers actually tell us about the distribution of density fluctuations? After all, \ii{non-Gaussianity} is a property of probability density functions (pdf) and thus it is important to understand exactly how $\fnl$ and $\gnl$ relate to the statistics of the density fluctuations. At leading order, $\fnl$ and $\gnl$ are proportional to the skewness and the excess kurtosis of the distribution of primordial density fluctuations \citep{desjacques}. But as we shall see in this paper, the knowledge of $\fnl$ and $\gnl$ is insufficient for the reconstruction the distribution of density  fluctuations at late times, as the corresponding pdf cannot be positive definite.

%As we shall see later, the exact relations between these parameters and the skewness and kurtosis are in the form of multidimensional integrals which are tedious to evaluate. 

It has been said that characterising non-Gaussianity is akin to characterising a ``non-dog". Whilst this is true in the sense that there are infinite possibilities of probability distributions that are not Gaussian, it is  misleading because different types of non-Gaussianity \ii{can} be systematically characterised, for instance, by the deviations in the moments or cumulants from the Gaussian values. This is, in fact, the idea behind the Edgeworth expansion, which expresses a weakly non-Gaussian distribution as the Gaussian distribution multiplied by a Taylor series consisting of cumulants (see \cite{blinnikov} for a review).

There have been a number of works that use the Edgeworth series to study the  distribution of density fluctuations \citep{bernardeau,juszkiewicz,amendola,loverde}. However, these works invariably truncate the Edgeworth expansion at just a few terms. In calculations of the abundance of massive clusters, the truncated series is often implicitly assumed to remain valid far out into the exponential tail of the distribution. In our opinion, it is very difficult to judge the validity of such calculations without establishing first the accuracy of the truncated Edgeworth series rigorously. In this paper, we shall address this issue by quantifying how sensitive the abundances of rare objects are to changes in the truncation order.

Many previous applications of the Edgeworth series also faced with the problem that the resulting pdf is negative in some region. This is often attributed to fact that there are an insufficient number of terms in the Edgeworth series. The conditions needed for the truncated Edgeworth expansions to be non-negative have been addressed in simple cases by a few works in the statistical literature \citep{draper,balitskaya,jondeau}. In this paper, we shall investigate this problem numerically and show that the positivity of the pdf generally depends not only on the number of terms in the series, but also on the available information on higher-order moments. We shall present some general guidelines on how the Edgeworth expansion could be used to model a non-negative non-Gaussian pdf.

Having developed the necessary background for proper use of the Edgeworth expansion, we shall apply it to study the effects of $\gnl$ on the number densities of massive clusters and large voids.

%In general, a given pdf and its representation in terms of a truncated Edgeworth series can behave quite differently. In particular, they contain potentially different statistical information, in the sense that the $n$-th moment of one may not be equal to that of the other. 

%Higher order statistical information such as the 5th moment are rarely discussed in the literature. This is surprising considering that there is no reason \ii{a priori} why they should be negligible, and that the technique for extracting such information from data is well understood\footnote{see \eg \cite{cappi,kurokawa} for calculations of higher moments using bootstrapped data from galaxy surveys.}. 

%See effect of $h\sub{NG}$ on the nonlinear mass function in \cite{maggiore3}.

\section{The primordial density fluctuations}

First, let us introduce the necessary parameters which will allow us to describe the distribution of density fluctuations statistically.

Let $\rho_c$, $\rho_b$, $\rho_r$, $\rho_\Lambda$ be the time-dependent energy densities of cold dark matter, baryons, radiation and dark energy. Let $\rho_m=\rho_c+\rho_b$. We define the density parameter for species $i$ as 
\be \Omega_i \equiv {\rho_i (z=0)\over \rho\sub{crit}},\ee
where $\rho\sub{crit}$ is the critical density defined by $\rho\sub{crit}\equiv 3H_0^2/8\pi G $. The Hubble constant, $H_0$, is parametrized by the usual formula $H_0\equiv100h \mbox{ km\ff s}^{-1}\mbox{Mpc}^{-1}$. Results from a range of astrophysical observations are consistent with $h\simeq 0.7$, $\Omega_c\simeq0.23$, $\Omega_b\simeq0.046$ and $\Omega_r\simeq 8.6\times10^{-5}$, with $\Omega_\Lambda=1-\Omega_m-\Omega_r$ [see \eg \cite{komatsu,lahav+}]. 
% don't use h unless necessary!!! 

The density fluctuation field, $\delta$, is defined as
\be \delta(\mb{x},t)\equiv {\rho_m(\mb{x},t)-\bkta{\rho_m(t)}\over\bkta{\rho_m(t)}},\ee
where $\bkta{\rho_m}$ is the mean matter energy density. In Fourier space, the density fluctuation field can be decomposed as
\be  \delta(\mb{x},t)=\int {d{\mb k}\over (2\pi)^3} \ff \delta(\mb{k},t) e^{i\mb{k}\cdot\mb{x}}.\ee As we shall be dealing mainly with observables measured at the present time, $t_0$, we simply write $\delta(\mb{k})$ to mean $\delta(\mb{k},t_0)$.

The gravitational Newtonian potential $\Phi$ is related to the density fluctuation by the cosmological Poisson equation. For a Fourier mode $\mb{k}$, this reads
\be \delta(\mb{k}) =  {2\over3\Omega_m }\bkt{k\over H_0}^2 \Phi(\mb{k}).\ee

Statistical information on $\delta(\mb{x})$ can be deduced from that of $\delta(\mb{k})$. The finite resolution of any observation, however, means that we can only empirically obtain information on the \ii{smoothed} moments of the distribution. More precisely, given a length scale $R$, the smoothed density field, $\delta_R$, observed today is given by 
\be \delta_R(\mb{k})=W(kR)D(0)T(k)\delta(\mb{k}),\ee
%where $D$ is the linear growth factor (which we calculate using the fitting formula of \cite{carroll,lahav}), 
where $k=|\mb{k}|$ and  $D(0)\approx 0.76$ is the linear growth factor evaluated at $z=0$. We choose $W$ to be the spherical top-hat function of radius $R$. In Fourier space, we have 
\be W(kR)=3\bkts{{\sin(kR)\over (kR)^3}-{\cos(kR)\over (kR)^2}}.\ee
It is also useful to define the mass of matter enclosed by the top-hat window as 
\be M\equiv {4\over3}\pi R^3\rho_m\approx 1.16\times10^{12}\bkt{{R\over h^{-1}\mbox{Mpc}}}^3 \ff h^{-1}M_\sun.\ee
We follow the approach outlined in \cite{weinberg} and use the transfer function $T$ of Dicus 
\begin{align} 
T(x)={\ln[1+(0.124x)^2]\over (0.124x)^2}\bkts{1+(1.257x)^2+(0.4452x)^4+(0.2197x)^6 \over1+(1.606x)^2+(0.8568x)^4+(0.3927x)^6  }^{1/2}.
%x&\equiv{\sqrt{2}k\over k\sub{eq}},
\end{align}
%where $k\sub{eq}$ is the wave-number of the Fourier mode that entered the Hubble radius at the epoch of matter-radiation equality. Roughly 
%\be k\sub{eq}\simeq[13.6 \mbox{Mpc}]^{-1} \Omega_mh^2.\ee
In addition, we also incorporate the baryonic correction of \cite{eisenstein}, whereby the transfer function is evaluated at 
\be x\sub{EH}={k\Omega_r^{1/2}\over H_0\Omega_m}\bkts{\alpha+{{1-\alpha}\over{1+(0.43ks)^4}}}^{-1},\lab{EH}\ee
% this is the 'effective' fit, ie. no oscillations - curve goes through mean of oscillations...  but still better than bbks or anything else.
% see eisenstein hu eq 30/31.. note that in that paper Omega_0 is actually Omega_m
with 
$$\alpha=1-0.328\ln(431\Omega_mh^2){\Omega_b\over\Omega_m}+0.38\ln(22.3\Omega_mh^2)\bkt{\Omega_b\over\Omega_m}^2,$$
and 
$$s={44.5\ln(9.83/\Omega_m h^2) \over \sqrt{1+10(\Omega_b h^2)^{3/4}}} \ff \mbox{Mpc}.$$

% follow Weinberg pg 408: this may affect factors of (2pi)^n in f_nl later on.. note that weinberg has different convention from lyth/liddle

The matter power spectrum, $P(k)$, can be defined via the two-point correlation function in Fourier space as
\be \bkta{\delta(\mb{k_1}),\delta(\mb{k_2})}=(2\pi)^3\delta_D(\mb{k_1}+\mb{k_2})P(k),\ee
where $\delta_D$ is the 3-dimensional Dirac delta function. In linear perturbation theory, it is usually assumed that inflation laid down an initial  spectrum of the form $k^{n_s}$, where $n_s$ is the scalar spectral index (assumed to be 0.96 in this work). Physical processes which evolve $P(k)$ through the various cosmological epochs can simply be condensed into the equation
\be P(k)\propto P_\phi(k)T^2(k),\ee
where $P_\phi(k)\propto k^{n_s-4}$. It is also common to define the dimensionless power spectrum $\mc{P}(k)$ as
\be \mc{P}(k)\equiv {k^3\over 2\pi^2} P_\phi(k)\propto \bkt{k\over H_0}^{n_s-1}.\ee
\no Consequently, the variance of density fluctuations smoothed on scale $R$ can be written as 
\be \sigma^2_R= \int_0^\infty {dk\over k} \ff A^2(k) \mc{P}(k),\lab{vari}\ee 
where 
\be A(k)= {2\over3\Omega_m }\bkt{k\over H_0}^2  W(kR) D(0) T(x\sub{EH}).\ee
In our numerical work, we shall normalise $\mc{P}(k)$ so that 
\be \sigma_8\equiv\sigma(R=8h^{-1}\mbox{Mpc})=0.8.\lab{sig8}\ee

%Given a set of unsmoothed cumulants $S_n$ ($n\geq3$), we model their scale dependence by the scaling relation
%\be S_n(R) = {B_n\over \sigma^{\alpha_n}_R}, \lab{model}\ee
%where the scaling index $\alpha$ determines the physical scales on which \ng manifests. In \cite{gaz2}, the scaling with $\alpha_n=n-2$ was presented by motivation of dimensional considerations. We will see later that this is a reasonably accurate approximation for the non-Gaussianity of `local' type parametrized by $\fnl$, but less so $\gnl$ and higher-order local non-Gaussianity.

%We model the \ng in the primordial pdf via $B_n$, which are assigned typically small values $\ll 1$. This allows us to isolate effects of the $n$th moment in the primordial distribution on observables. Later on, we also study the effects of changing the scaling via $\alpha(n)$, in which case $\fnl$ fails to be a robust description for non-Gaussianity.

\section{Statistical information in $\fnl$ and $\gnl$}\lab{fnlgnl}

The most widely studied type of \ng is the `local' type parametrized, at lowest orders, by $\fnl$ and $\gnl$, which are the coefficients in the Taylor expansion of the non-linear Newtonian potential, $\Phi$, in terms of the linear, Gaussian field, $\phi$,
\be \Phi(\mb{x})= \phi(\mb{x})+\fnl\bkt{\phi^2(\mb{x})-\bkta{\phi^2}}+\gnl\phi^3(\mb{x})+\ldots.\lab{expand}\ee

This form of non-Gaussianity arises in simple models of single and multi-field inflation \citep{bartolo,rigopoulos,byrnes} as well as some curvaton models \citep{bartolo2,sasaki}. In this work, we shall assume non-Gaussianity only of this form. In general, it is possible that non-Gaussianity may be non-local. Mechanisms such as DBI inflation \citep{alishahiha} or inflation with a non-standard Lagrangian \citep{arkani-hamed,chen3} are known to generate primarily non-local non-Gausssianity. We comment on these possibilities later, but leave a full investigation for future work.

We adopt the `large-scale-structure' convention in which $\Phi$ is extrapolated to $z=0$. We also take $\fnl$ and $\gnl$ to be constant, although it is conceivable that they may be scale dependent (see \cite{sefusatti,cayon} for  constraints on the `running' of $\fnl$). In this section, we investigate how this form of \ng is related to the reduced cumulants, $S_n$, defined by
\be S_{n}(R) \equiv  {\bkta{\delta_R^n}_c\over \sigma_R^{2n-2}},\lab{redcum}\ee
where $\bkta{\delta_R^n}_c$ is the $n$th cumulant. For a distribution with zero mean, the relationships between the first few cumulants and moments are
\be \bkta{\delta_R}_c &=& 0, \qquad \bkta{\delta_R^2}_c= \sigma_R^2, \nn\\
\bkta{\delta_R^3}_c &=& \bkta{\delta_R^3}, \quad \bkta{\delta_R^4}_c= \bkta{\delta_R^4}-3\sigma_R^4.\lab{cum}\ee

Throughout this work we shall often make references to the \ii{skewness} and \ii{kurtosis}, which are defined respectively as $\bkta{\delta_R^3}/\sigma_R^3$ and $\bkta{\delta_R^4}/\sigma_R^4$. The \ii{excess kurtosis} is defined as $\bkta{\delta_R^4}/\sigma_R^4-3$, with 3 being the kurtosis of the Gaussian distribution.

\bbb

\subsection{$f\sub{NL}$}

\begin{figure}
\centering
\epsfig{file=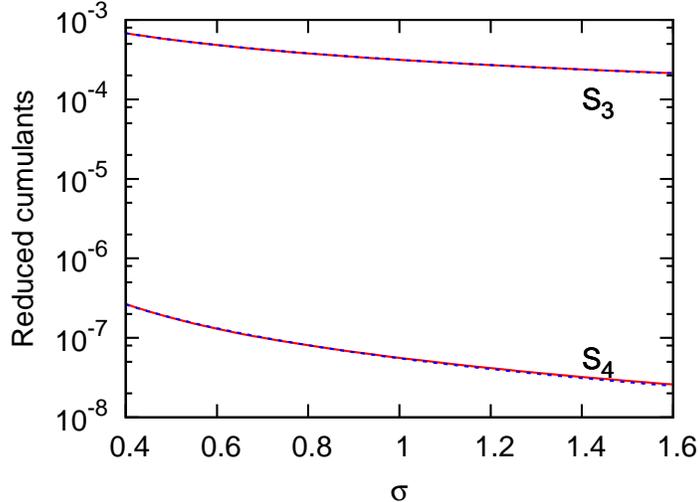, width= 7cm, angle = -90}
\caption{ Solid lines show the reduced cumulants $ S_3$ and $ S_4$ for $\fnl=1$ and $\gnl=1$ as a function $\sigma(M)$, with mass $M$ ranging between $10^{13}-10^{16}\,h^{-1}M_\sun$. Higher values of $\fnl$ and $\gnl$ scale multiplicatively. The overlapping dashed lines show the fitting formulae given by equations \re{scale1} and \re{scale2}.}
\label{figcum}
\end{figure}

At leading order, $\fnl$ is related to the skewness via the relation derived in \cite{desjacques}
\be \sigma^4  S_3(R) = \fnl \int_0^\infty \!\!{dk_1\over k_1}\ff A(k_1) \mc{P}(k_1) \int_0^\infty \!\!{dk_2\over k_2} \ff  A(k_2) \mc{P}(k_2) \int_{-1}^1\!\! d\mu \ff A(k_3)\bkts{1+2{P_\phi(k_3)\over P_\phi(k_2)}}, \lab{bigfnl}\ee
where $k_3^2=k_1^2+k_2^2+2\mu k_1k_2$. Figure \ref{figcum} (upper curve) shows the cumulant $S_3$ as a function of $\sigma_R$ in the mass range $10^{13}-10^{16}\ff h^{-1}$Mpc. On these scales, the weak scale-dependence in $S_3$ can be accurately fitted by a simple formula
\be  S_3 \simeq {3.15\times 10^{-4}\times\fnl\over \sigma_R^{0.838}},\lab{scale1}\ee

\no with sub-percent accuracy. This fitting formula offers an easy way to calculate the observational signatures of $\fnl$ without resorting to the integrals in \re{bigfnl}.

%%%%%%%%%%%%%%%%%%%%%%%%%%

%%%%%%%%%%%%%%%%%%%%%%%%%%
%%%%%%%%%%%%%%%%%%%%%%%%%%
%%%%%%%%%%%%%%%%%%%%%%%%%%

\subsection{$g\sub{NL}$}

If $\fnl=0$, we can similarly derive the leading-order relation between $\gnl$ and the excess kurtosis 

\be\sigma^6S_4(R)&=& {3\over32\pi^3}\gnl\bkt{\prod_{i=1}^{3}\int_0^\infty{dk_i\over k_i}\ff A(k_i) \mc{P}(k_i) \int_{-1}^1\!\! d\mu_i  \int_{0}^{2\pi}\!\! d\phi_i} A(k_{4})\bkts{1+3{P_\phi(k_4)\over P_\phi(k_3)}}, \lab{biggnl}\ee

where 
\be k_{4}&\equiv&\bkt{k_1^2+k_2^2+k_3^2+2 k_1k_2\Theta_{12}+2k_2k_3\Theta_{23}+2k_1k_3\Theta_{13}}^{1/2},\\
\Theta_{ij}&\equiv&\bkts{(1-\mu_i^2)(1-\mu_j^2)\cos(\phi_i-\phi_j)+\mu_i\mu_j}^{1/2}.
\ee 

The result of this integration is shown as the lower curve in figure \ref{figcum}. As before, we found a fitting formula
\be  S_4 \simeq {5.53\times 10^{-8}\times\gnl\over {\sigma_R^{1.70}} },\lab{scale2}\ee
accurate on the same mass scale. 

In summary, we can easily emulate the effects of $\fnl$ and $\gnl$ using the fitting formulae \re{scale1} and \re{scale2} without having to compute the multidimensional integrals \re{bigfnl} and \re{biggnl}. Of course, these formulae depend on the choice the primordial power spectrum as well as the window and transfer functions. The fitting formulae are not expected to be very sensitive to the changes in any one these ingredients.

\bbb
\section{Non-Gaussianity based on $\fnl$ and $\gnl$ only}\lab{sectionrose}

Much effort has been placed into using the parameters $\fnl$ and $\gnl$ to specify the deviation of the primordial distribution from Gaussianity. As shown in the previous section, these parameters correspond, at leading order, to  deviations from Gaussianity in the 3rd and 4th moments of the distribution. It is important to examine if one can consistently parametrize a non-Gaussian distribution in this way without having to worry about deviations in the higher-order moments. We shall demonstrate that this cannot be the case.

The objective here is to reconstruct the pdf of density fluctuations given a sequence of moments $\{\alpha_n\equiv\bkta{\delta_R^n},n=0,1,2\ldots\}$, which, in practice, can be estimated from galaxy-survey data \citep{cappi,kurokawa}. This is the classic \ii{moment problem} which has been studied in great detail,  beginning with the pioneering work of Stieltjes in 1894 and Hamburger in 1920 (see \cite{kjeldsen} for a historical review). In general, there is no guarantee that the resulting pdf will be non-negative, or, indeed, that a solution exists at all. A useful theorem regarding the existence of a solution to the \ii{Hamburger moment problem}, \iee when the pdf is defined on $(-\infty,\infty)$, is the following:

\newtheorem*{thmm}{Theorem}

\begin{thmm}[\ii{Existence of solution to the Hamburger moment problem}] The sequence $\{\alpha_n, n=0,1,2\ldots\}$ corresponds to moments of a non-negative pdf if and only if the determinants 

\be D_n = \begin{vmatrix} \alpha_0 & \alpha_1 &\alpha_2 & \ldots & \alpha_n \\
\alpha_1 & \alpha_2 &\alpha_3 & \ldots & \alpha_{n+1}\\
\alpha_2 & \alpha_3 &\alpha_4 & \ldots & \alpha_{n+2}\\
\vdots & \vdots & \vdots & \vdots & \vdots \\
\alpha_n & \alpha_{n+1} &\alpha_{n+2} & \ldots & \alpha_{2n}   
\end{vmatrix}, \ff\ff\ff n=0,1,2\ldots\lab{determinant}\ee
are all non-negative.
\end{thmm}

See \cite{shohat} or \cite{akheizer} for proof.

\sss

Let us consider the standardised distribution of $x=\delta_R/\sigma_R$, so that $\alpha_0=1, \alpha_1=0$ and $\alpha_2=1$. Let us also suppose (as is implicit in some previous works) that non-Gaussianity weakly manifests in the skewness and kurtosis only (therefore $\alpha_3$ is close to 0 and $\alpha_4$ is close to 3). Higher moments are taken to be identical to those of the normal distribution
\be \alpha_{n}= {\begin{cases} (n-1)!! =1\cdot3\cdot5\cdots (n-1), &\ff n \mbox{ even}\\ 0,  &\ff n \mbox{ odd}\end{cases}}\quad (n\geq5).\ee
%where the double factorial denotes the product of every odd number from $1$ to $2n-1$ inclusive.

The expressions for $D_n$ up to $n=6$ are given below. For convenience, we write $\sfs=\alpha_3$ (skewness) and $\sfk=\alpha_4-3$ (excess kurtosis).
\be
D_3 &=& \sfk-\sfs^2+2, \nn\\
D_4 &=& -\sfk^3-8\sfk^2-6\sfk-3\sfk\sfs^2-24\sfs^2+\sfs^4 +12,\nn\\
D_5 &=& \sfk^5+15\sfk^4-60\sfk^3-750\sfk^2+360\sfk-45\sfk\sfs^2+45\sfk^2\sfs^2-1890\sfs^2+105\sfs^4+288,\nn\\
D_6 &=& 945\sfk^5+11025\sfk^4-80100\sfk^3-324000\sfk^2-43200\sfk-132300\sfk\sfs^2+31500\sfk^2\sfs^2-518400\sfs^2+99225\sfs^4+34560.\nn
\ee
Note that for the Gaussian distribution ($\sfs=\sfk=0$), these $D_n$'s are all positive as expected.
%%%%%%%%%%%%%%%%%%%%%%%%%%

\begin{figure}
\centering
\epsfig{file=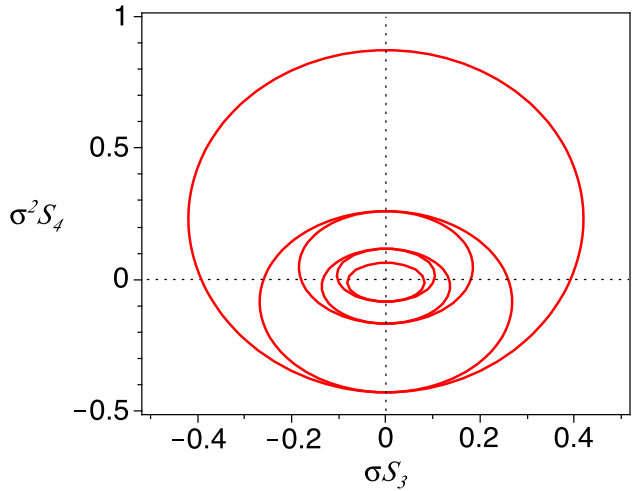, width= 8.5cm}
\caption{Regions in which the determinant $D_n\geq0$ [see theorem \re{determinant}]. The horizontal axis shows $\sigma S_3$ (skewness) and the vertical axis shows $\sigma^2 S_4$ (excess kurtosis). The regions correspond to $n=5$ (outermost ellipse) to $n=10$ (innermost ellipse). The regions are nested, co-tangential and converge towards the origin as $n\rightarrow\infty$. The latter fact implies that there is no well-defined non-Gaussian pdf that can be described by deviations in the skewness and kurtosis alone.}
\label{rose}
\end{figure} 

%%%%%%%%%%%%%%%%%%%%%%%%%%
%%%%%%%%%%%%%%%%%%%%%%%%%%

The condition $D_n\geq0$ always describes a closed region in the $(\sfs,\sfk)$ plane containing the origin and bounded by the curve $D_n=0$. Figure \re{rose} shows these regions for $n=5$ (outermost ellipse) to $n=10$ (innermost ellipse). To make a connection with later sections, we have labelled the axes as $(\sigma S_3, \sigma^2 S_4)$, where
\be \sigma S_3 = \sfs,\quad \sigma^2 S_4 = \sfk,\ee
as can be easily shown using relation \re{redcum}-\re{cum} (to avoid cluttering we sometimes write $\sigma$ to mean $\sigma_R$). As $n$ increases, the region corresponding to $D_n\geq0$ becomes smaller. Interestingly, the regions for any two consecutive values of $n$ are nested and co-tangential. One can continue inductively this way to find that as $n\rightarrow\infty$, the ellipses converge to the origin, implying that there is no room for any deviation from Gaussianity. We conclude that deviations in the skewness and kurtosis alone cannot consistently parametrize a non-Gaussian pdf. 

The upshot of all this is that $\fnl$ and $\gnl$ by themselves cannot completely describe a non-Gaussian pdf. Information on higher-order correlation must be available for the pdf to be well defined.

\section{The Edgeworth Expansion}

As described in the Introduction, the Edgeworth expansion is a convenient way to express a weakly non-Gaussian pdf as a series comprising its cumulants. Suppose that we only have estimates on $\fnl$ and $\gnl$, and no higher-order non-Gaussianity. The result of the previous section shows that the resulting pdf cannot be non-negative.

Nevertheless, this result only holds if we use an infinite number of cumulants in the reconstruction of the pdf. This is equivalent to having an infinite number terms in the Edgeworth expansion. In numerical implementations, however, one truncates the Edgeworth expansion after a finite number of terms. As we will see shortly, it now becomes possible to describe an entirely non-negative pdf with only $\fnl$ and $\gnl$, circumventing the result of the previous section. The disadvantage of the truncation is, of course, that the cumulants of the reconstructed pdf may not correspond exactly to those of the actual pdf, and thus the statistical significance of $\fnl$ and $\gnl$ is somewhat weakened. For very short series of just a few terms, the interpretations of $\fnl$ as skewness and $\gnl$ as excess kurtosis are especially dubious.

Given information on a finite number of cumulants, we shall investigate the sensitivity of the resulting pdf to the number of terms in the Edgeworth expansion. This sensitivity has been alluded to by several works in the literature \citep{juszkiewicz,loverde,desjacques}, though we believe that our analysis goes beyond those works. In particular, we shall argue that the truncated series cannot be used to deduce results for negative $\gnl$, unless some higher-order non-Gaussianities are known.

\subsection{The Petrov development}

In this paper, we shall be using the form of the Edgeworth series given by \cite{petrov}, who gave a method of calculating the Edgeworth series to arbitrarily high order. Given a non-Gaussian pdf with zero mean and variance $\sigma_R^2$, we can express its deviation from Gaussianity as a product of the normal distribution and a Taylor series in $\sigma_R$:
\be p(\delta_R) = N(\delta_R)\bkts{1+\sum_{s=1}^{\infty}\sigma_R^sE_s\!\!\bkt{\delta_R\over\sigma_R}},\lab{edgeworth}\ee
where $N(\delta_R)$ is the normal distribution
\be N(\delta_R)={1\over \sigma_R\sqrt{2\pi}}\exp\bkt{-\delta_R^2\over 2\sigma_R^2},\ee
and the coefficients $E_s$ in the Taylor series are given by
\be
E_s\bkt{\nu}&=&\!\sum_{\{k_m\}}\bkts{H_{s+2r}\!\bkt{\nu}\prod_{m=1}^{s}{1\over k_m!}\bkt{S_{m+2}\over (m+2)!}^{k_m}\!},\ff
\lab{coeff}\\
\mbox{where}\quad\nu&\equiv&{\delta_R\over\sigma_R}.\nn\ee We now explain the various components of the coefficient \re{coeff}. Firstly, the sum is taken over all distinct sets of non-negative integers $\{k_m\}_{m=1}^{s}$ satisfying the Diophantine equation
\be k_1+2k_2+\ldots +sk_s = s.\ee We also define
\be r\equiv k_1+k_2+\ldots +k_s.\ee
Next, the function $H_n(\nu)$ is the Hermite polynomial of degree $n$. They can be obtained by the Rodrigues' formula
\be H_n(\nu)= (-1)^n e^{\nu^2/2}\diff{^n}{\nu^n}\bkt{e^{-{\nu^2/2}}}.\ee For example, $H_0(\nu)=1$ and $H_1(\nu)=\nu$. Higher order polynomials can be easily obtained via the recurrence relation
\be H_{n+1}(\nu)=\nu H_n(\nu)-n H_{n-1}(\nu).\ee
Note that if $p(\delta_R)$ is Gaussian, the cumulants of order $\geq3$  vanish identically, and so do the expansion coefficients \re{coeff}, as one might expect.

\subsection{Validity of the truncated series}

The Edgeworth expansion takes, as input, a sequence of cumulants $\{S_n\}$ which are combined with polynomials of various degrees up to order $N$. Therefore, when using the Edgeworth expansion, there are two factors which will determine its accuracy, namely 1) the number of available cumulants, and 2) the order $N$. These two issues are separate in the sense that it is possible to expand the Edgeworth series to arbitrarily high order given a limited number of cumulants. Both these issues must be analysed to properly monitor the sources of error.

\subsubsection{Linear truncation}
When the Edgeworth series \re{edgeworth} is truncated at linear order in $\sigma_R$, the resulting pdf is given by

\be p(\nu) = N(\nu)\bkts{1+{\sigma_R  S_3 \over 6}(\nu^3-3\nu)},\ee
where $\nu\equiv \delta_R/\sigma_R$ as before. Observe that if $ S_3>0$, a sufficiently large negative $\nu$ gives $p(\nu)<0$ and, similarly, if $S_3<0$, a sufficiently large positive $\nu$ gives $p(\nu)<0$. For instance, if $|S_3|=0.1$, $p(\nu)$ becomes negative as early as $|\nu|\simeq3$. This implies that a linear truncation of the Edgeworth series is highly suspect and is certainly not suitable for calculating, for instance, the mass function whereby high values of density fluctuations are involved.

\subsubsection{Quadratic truncation}

\begin{figure}
\centering
\epsfig{file=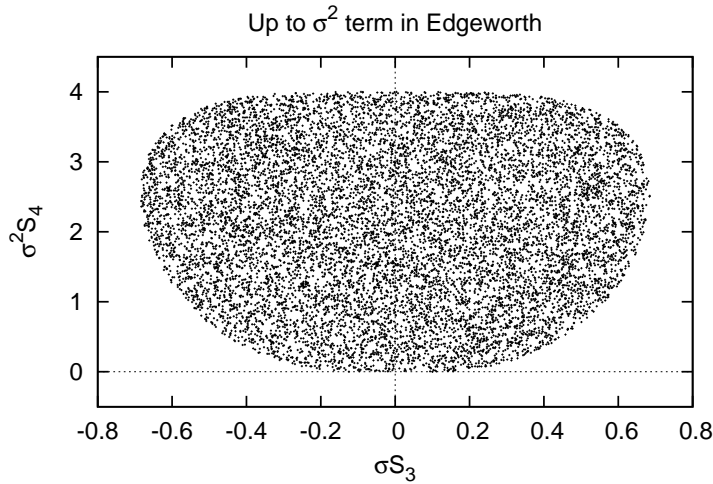, width= 7cm, angle = -90}
\caption{Validity of the quadratic Edgeworth expansion \re{quadratic}. The shaded region corresponds to the combination of $ S_3$ and $ S_4$ for which there exists a non-negative pdf over the entire real line. See the text for further discussions and proof of the bound $0\leq\sigma^2 S_4\leq4$.}
\label{scatter}
\end{figure}

The Edgeworth series truncated at quadratic order in $\sigma_R$ yields
% N(x)={1\over\sqrt{2\pi}}e^{-x^2/2}
\be p(\nu) = N(\nu)\bkts{1+{\sigma_R  S_3 \over 6}H_3(\nu)+\sigma_R^2\bkt{{ S_4\over24}H_4(\nu)+{ S_3^2\over72}H_6(\nu)}}.\lab{quadratic}\ee
We wish to determine the combination of $ S_3$ and $ S_4$ such that $p(\nu)$ is non-negative. The numerical evaluation of $p(\nu)$ over a grid of $S_3$ and $ S_4$ is shown in figure \ref{scatter}. In the figure, we mark points for which $p(\nu)>0$ in the $(\sigma_R S_3,\sigma_R^2 S_4)$ plane, in the domain $\nu\in[-20,20]$. This reveals a closed region in which $p(\nu)>0$. In fact, the bounding envelope can be found analytically by setting $p(\nu)=p^\pr(\nu)=0$, but the resulting equation has a very complicated parametric form which we shall not show here. For details of this technique see \cite{jondeau}.

%%%%%%%%%%%%%%%%%%%%%%%%%%

%%%%%%%%%%%%%%%%%%%%%%%%%%
%%%%%%%%%%%%%%%%%%%%%%%%%%

A curious feature of figure \ref{scatter} is that the excess kurtosis, $\sigma^2 S_4$, is limited to a small, non-negative range. We can prove this as follows. Setting $S_3=0$, the quadratic series can be written as 
\be {p(\nu)\over N(\nu)}= 1+{\sigma_R^2S_4\over 24}\bkts{(\nu^2-3)^2-6}.\ee
This expression clearly achieves the minimum when $\nu^2-3=0$. Requiring the minimum to be non-negative establishes the upper bound $\sigma^2S_4\leq4$. Next, if $S_4<0$, the quartic expression is unbounded from below and so the pdf will be negative for some large $x$. Thus, we must have $0\leq\sigma^2 S_4\leq4$

The bound for $\sigma S_3$ is more difficult to establish and we shall not go into the detail here. We simply note since an analytic expression describing the shaded region  in figure \ref{scatter} exists, the region in fact represents combinations of $S_3$ and $S_4$ ($\fnl$ and $\gnl$) for which the pdf is non-negative on the entire real line and not just in [-20,20]. For higher-order truncations, it becomes increasingly difficult to find such a region, as expected given the conclusion in section \ref{sectionrose}

%%%%%%%%%%%%%%%%%%%%%%%%%%

\begin{figure}
\centering
\epsfig{file=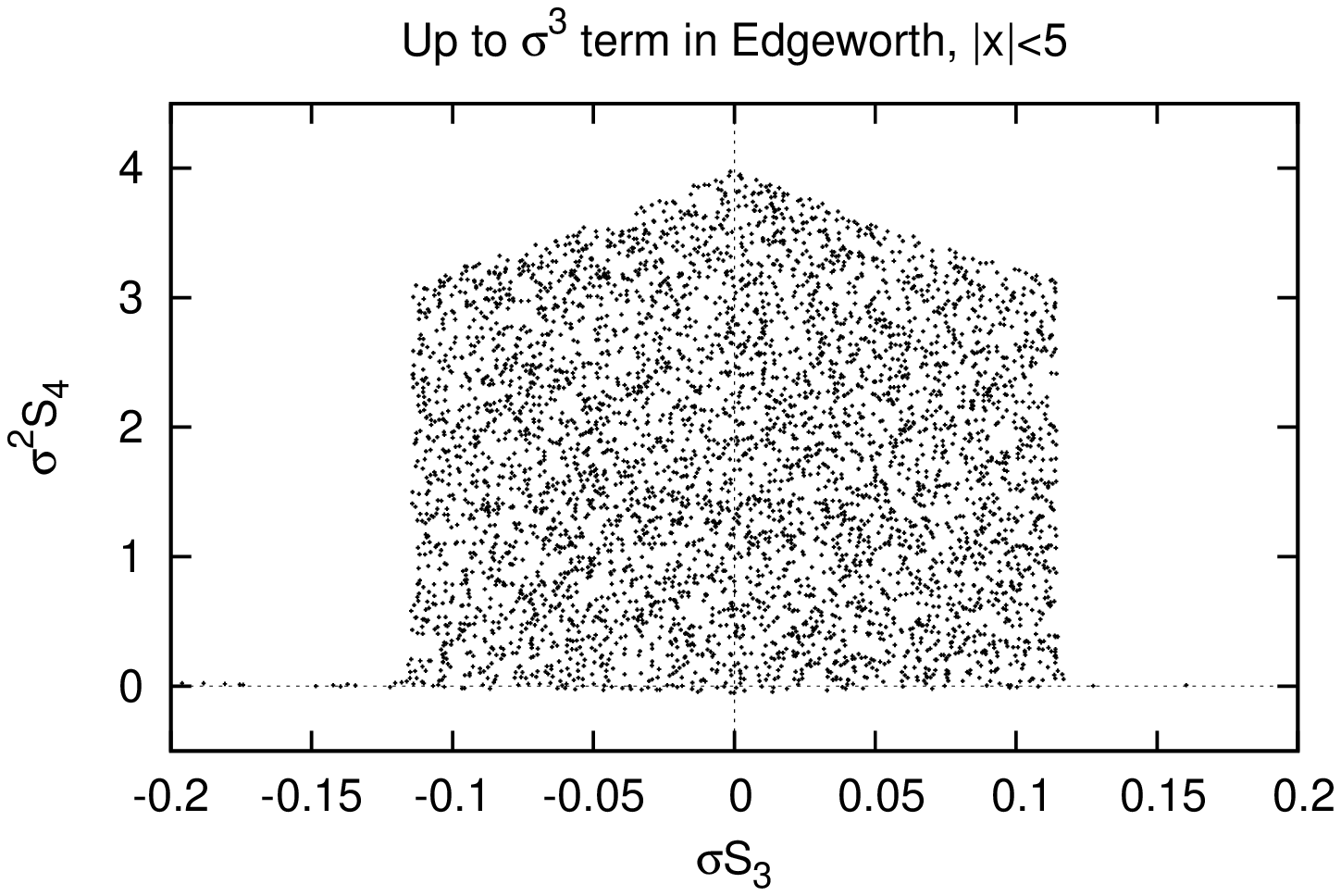, width= 5.1cm, angle = -90}\epsfig{file=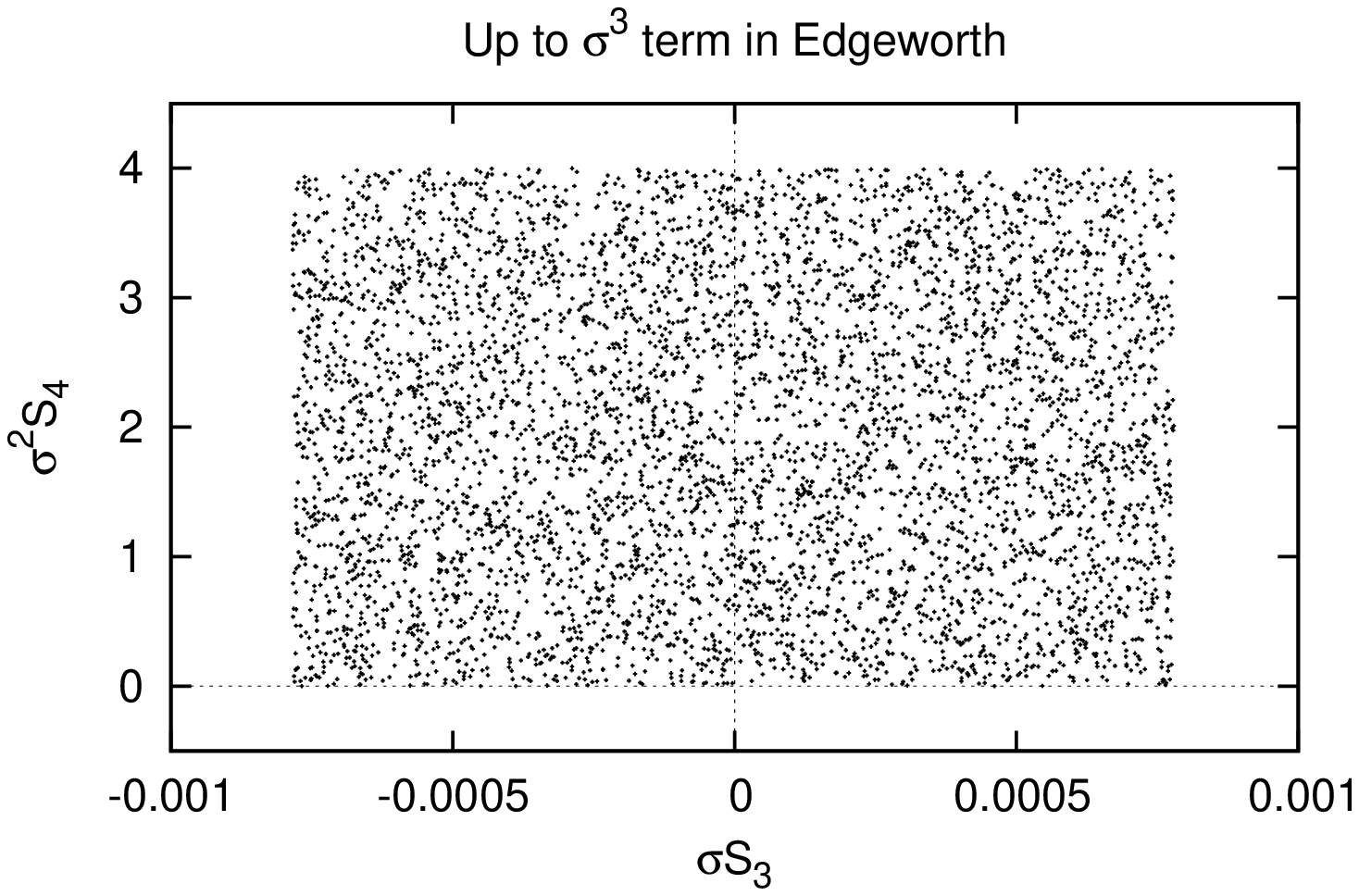, width= 5.1cm, angle = -90}

\epsfig{file=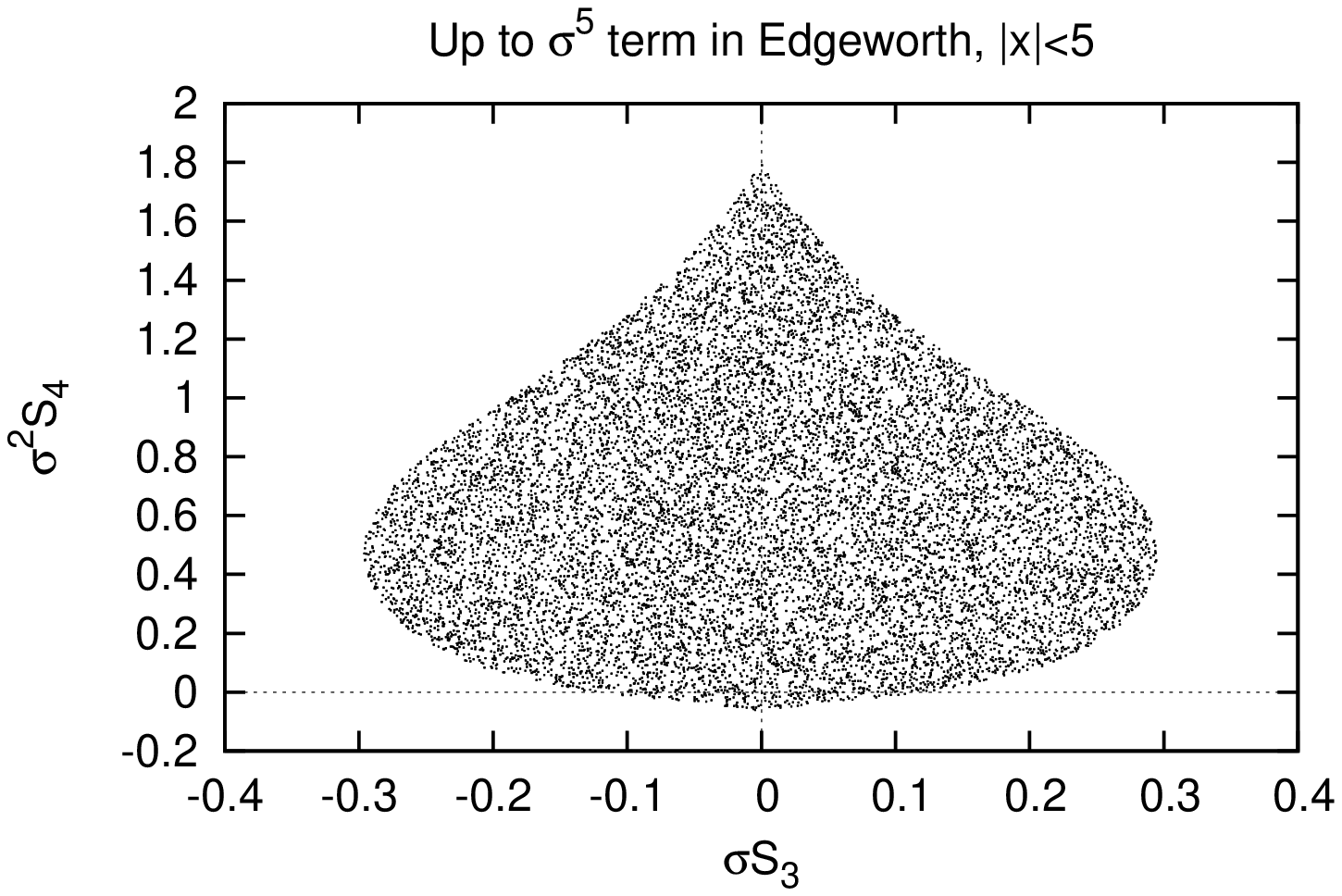, width= 5.1cm, angle = -90}\epsfig{file=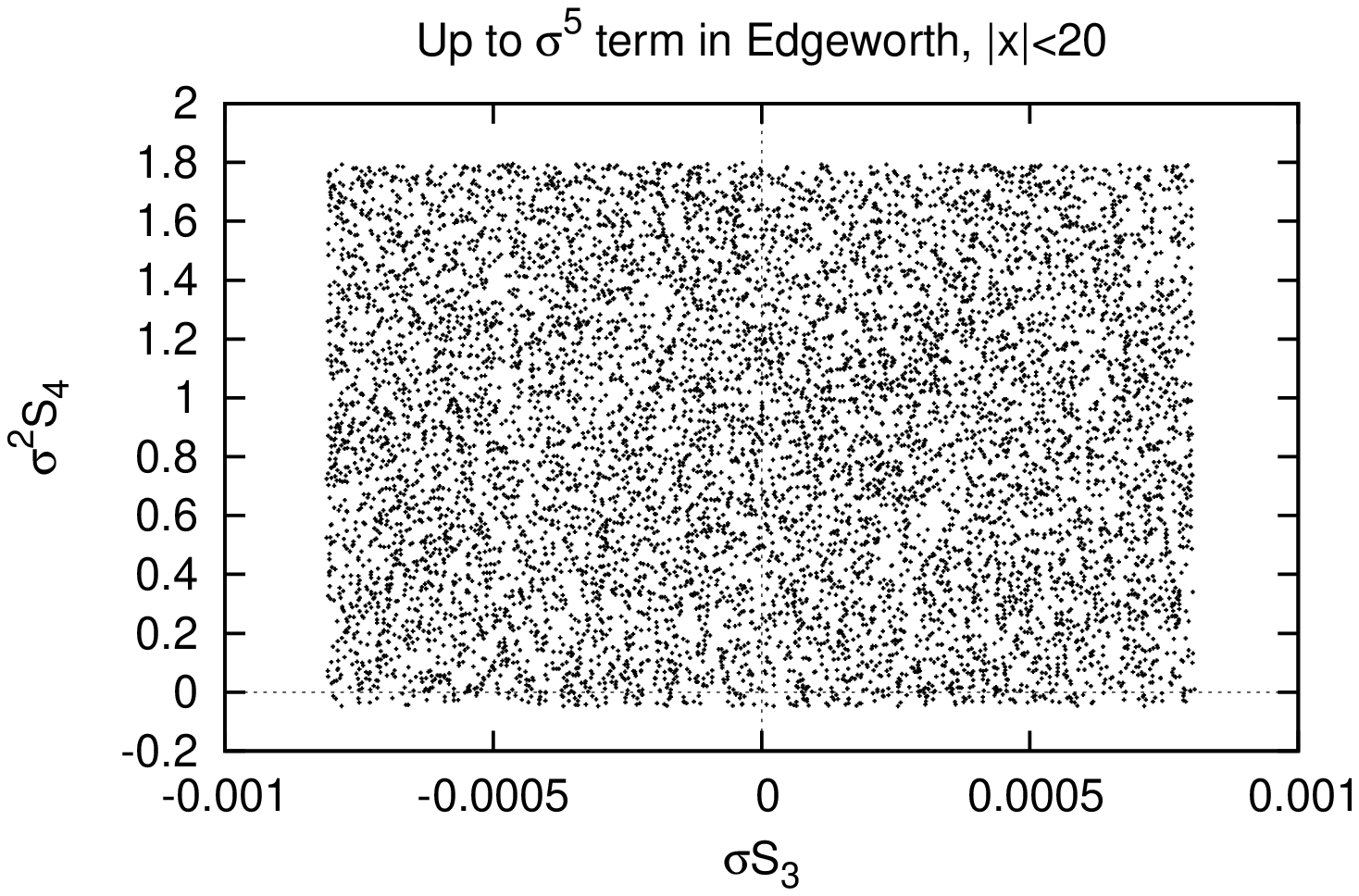, width= 5.1cm, angle = -90}

\epsfig{file=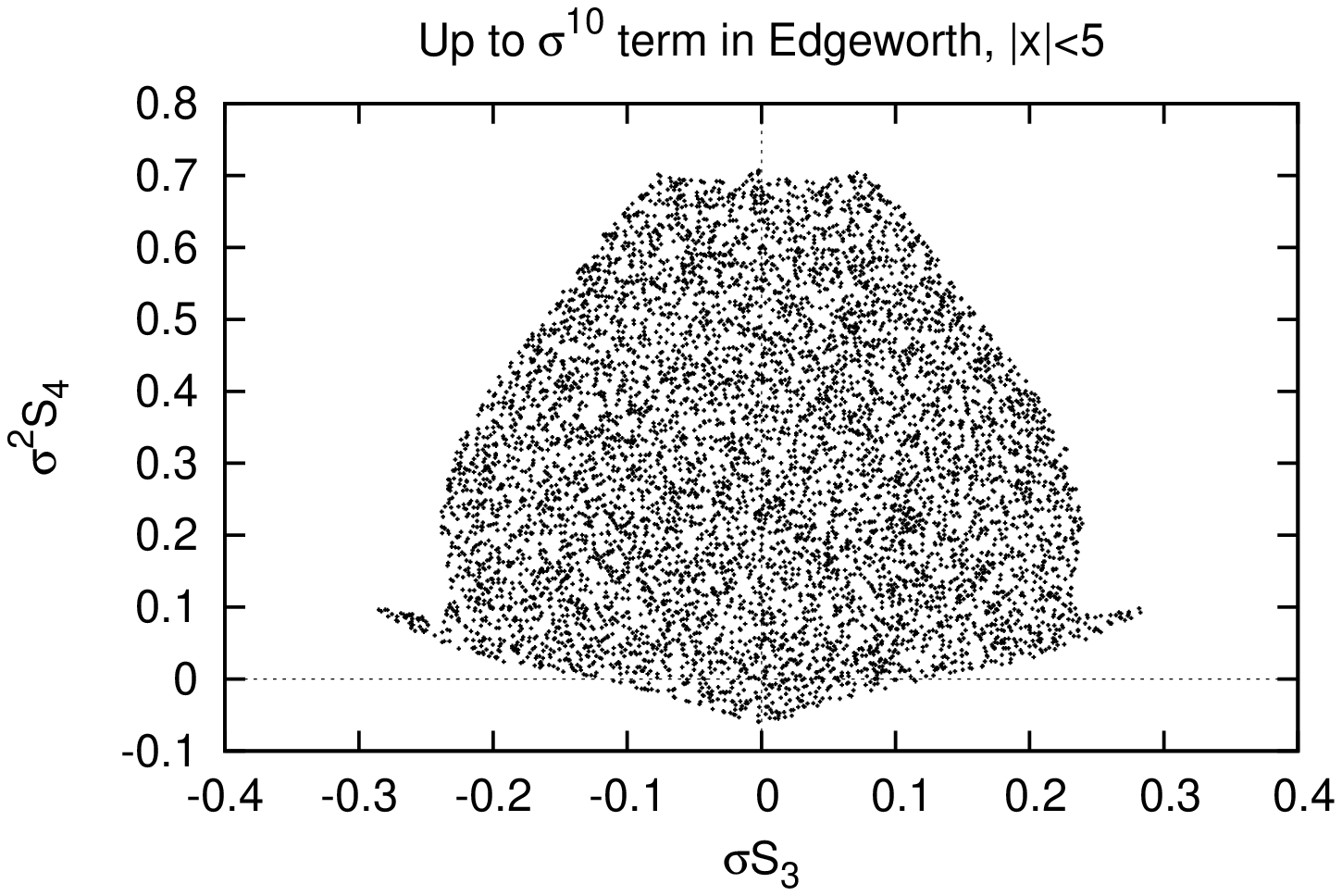, width= 5.1cm, angle = -90}\epsfig{file=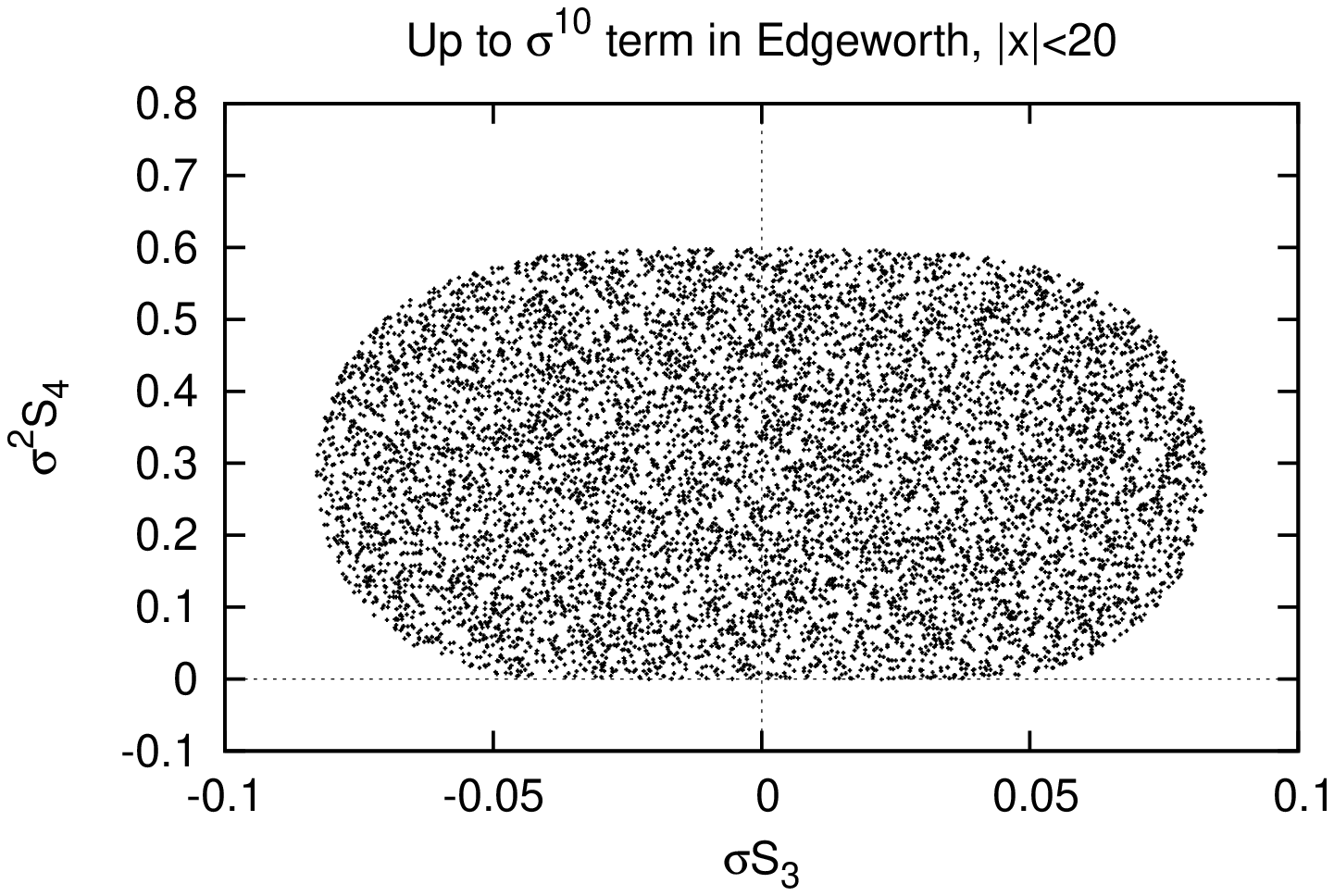, width= 5.1cm, angle = -90}

\epsfig{file=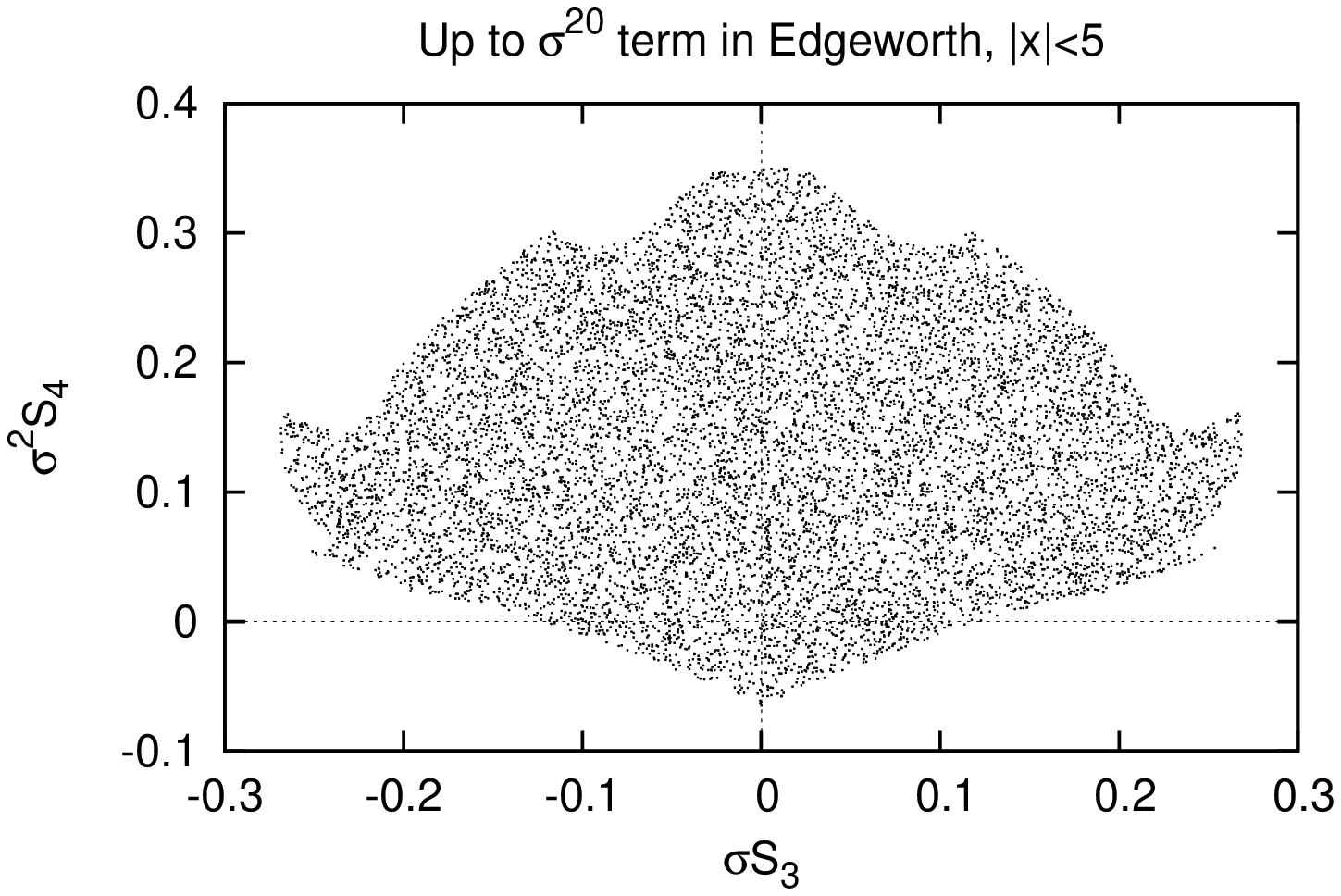, width= 5.1cm, angle = -90}\epsfig{file=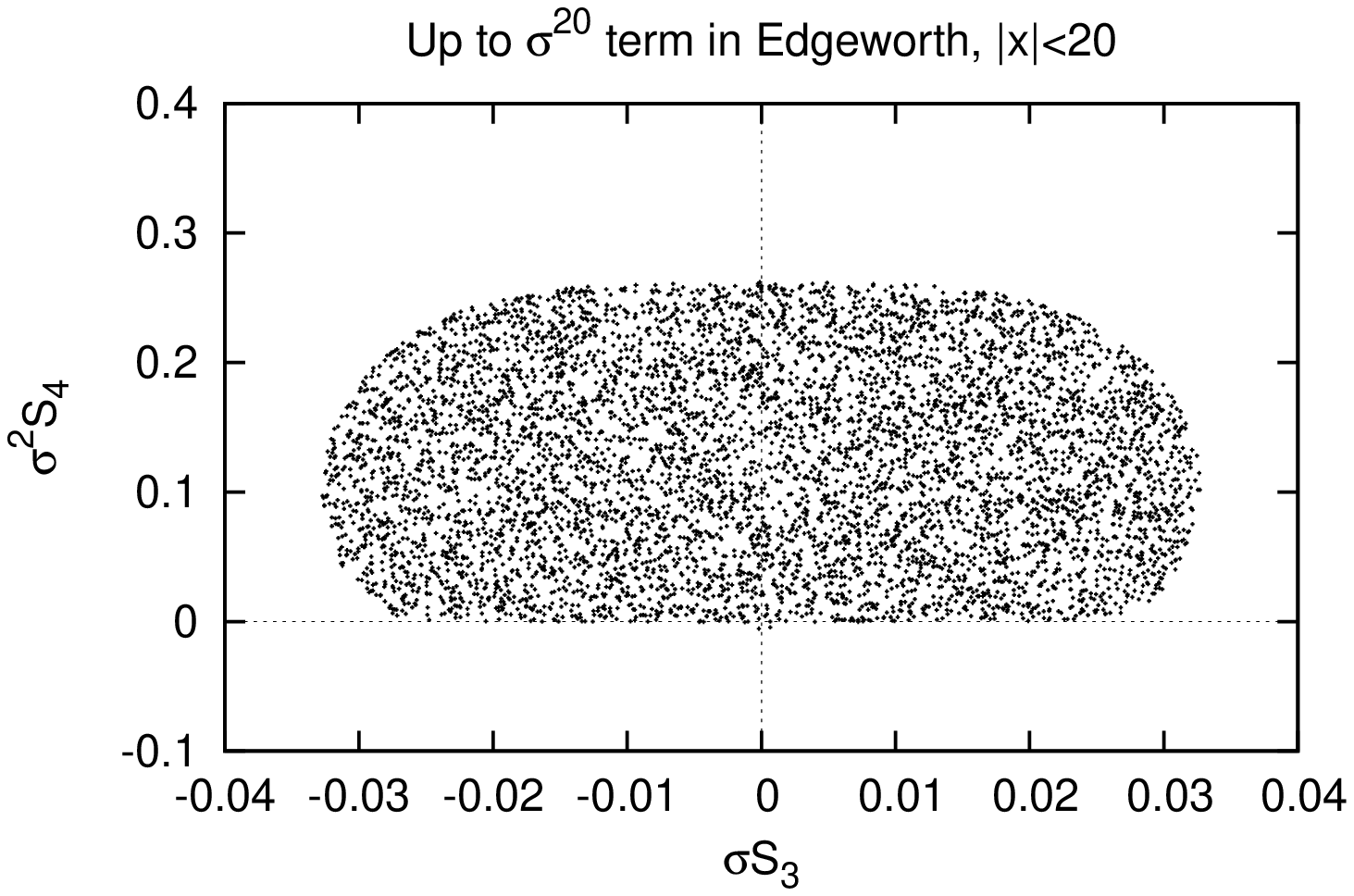, width= 5.1cm, angle = -90}

\caption{Validity of the Edgeworth expansion in the $(\sigma S_3,\sigma^2 S_4)$  plane for $\nu\equiv \delta/\sigma$ in the range $[-5,5]$ (column on the left) and $[-20,20]$ (right)  with the series is expanded up to (from top to bottom) $\sigma^3$, $\sigma^5$, $\sigma^{10}$ and $\sigma^{20}$. Only deviations in the skewness and the excess kurtosis are taken into account, with no higher-order non-Gaussianities. Points correspond to those combinations resulting in a non-negative pdf. See the text for further discussions on how the Edgeworth expansion may be properly used to produce non-negative pdfs. }
\label{scatterlots}
\end{figure} 

%%%%%%%%%%%%%%%%%%%%%%%%%%
%%%%%%%%%%%%%%%%%%%%%%%%%%

\subsubsection{Higher-order truncations}

%%%%%%%%%%%%%%%%%%%%%%%%%%

\begin{figure}
\centering
\epsfig{file=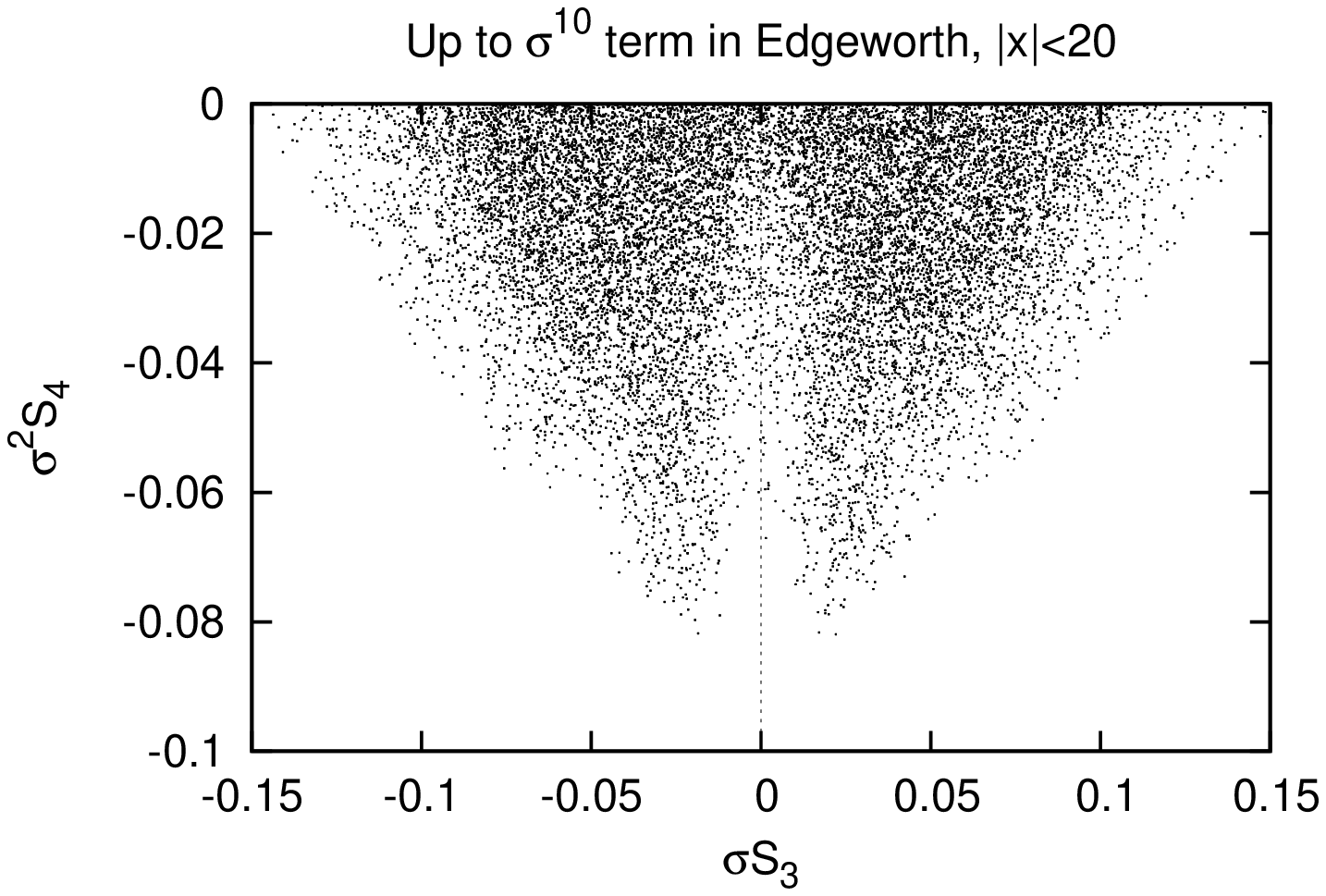, width= 6.4cm, angle = -90}
\caption{Including non-zero 6th and 8th cumulants ($S_6$ and $S_8$) opens up the region of validity of the truncated Edgeworth series to include models with $S_4<0$ (compare with figure \ref{scatterlots}, 3rd row, right column). Only points with $S_4<0$ are shown. For all these points, the 8th cumulant, $S_8$, is positive, as explained in the text.}
\label{s6}
\end{figure} 

%%%%%%%%%%%%%%%%%%%%%%%%%%
%%%%%%%%%%%%%%%%%%%%%%%%%%

Figure \ref{scatterlots} shows the same set of axes as figure \ref{scatter} with the Edgeworth series now expanded up to terms of order $\sigma^3$, $\sigma^5$, $\sigma^{10}$ and $\sigma^{20}$ (top row to bottom row). In producing these figures, we have set the rest of the cumulants to zero. This is roughly equivalent to parametrizing the non-Gaussianity by $\fnl$ and $\gnl$ only. Note that if $n$ is odd, the series up to $n$ terms performs significantly worse than one with even $n$. This is simply because odd (Hermite) polynomials are not positive definite, whereas even ones are, provided the coefficients are properly chosen. When scanning over a sufficiently large range of $\nu$, an odd-ordered Edgeworth expansion will not produce any well-defined pdf whatsoever. 

The sensitivity of the regions to the range of $\nu$ considered is clearly seen in the difference between the column on the left (in which $p(\nu)$ is only required to be non-negative for $|\nu|<5$) and on the right ($|\nu|<20$). As the range of $\nu$ increases, the cluster of points shrinks as it becomes increasingly difficult to find a closed region with $p(\nu)>0$. 

%This fact is crucial, for if it is generically true, then it implies that any attempt to model non-Gaussianity with skewness (or $\fnl$) \ii{alone} is immediately thwarted by the fact that the resulting pdf is ill-defined regardless of how many terms there are in the Edgeworth expansion.

Observe that for $\nu\in[-20,20]$, very few models with negative $S_4$ (\ie $\ff\gnl<0$) are produced. In fact, if the range of $\nu$ is sufficiently large, no models with negative $S_4$ are produced at all. A simple explanation for this is as follows. If the highest non-zero cumulant of a non-Gaussian distribution is $S_4$, then, for large $\nu$, the Edgeworth series expanded to $n$ terms is of order $S_4 \nu^n$. Hence, if $S_4<0$, a sufficiently large $\nu$ will render the expansion negative regardless of the value of $n$. 

Therefore, it is necessary that higher-order non-Gaussianities are taken into account when modelling a non-Gaussian distribution with $\gnl<0$. For instance, including nonzero cumulants $S_6$ and $S_8$ opens up the parameter space to those with $S_4<0$, as shown in figure \ref{s6}.

In summary, the Edgeworth series should be expanded up to even order in $\sigma_R$ to produce a well-defined pdf. The highest cumulant in that case is restricted to non-negative values. The Edgeworth expansion therefore can describe models with $\gnl<0$ if and only if cumulants of order at least $6$ or higher are included. If $\gnl=0$ and non-Gaussianity is parametrized by $\fnl$ only, the Edgeworth expansion is odd-ordered and the resulting pdf is not well-defined. 

Although we have assumed that non-Gaussianity is characterised purely by the `local' $\fnl$ and $\gnl$ parameters, the results in this section (as summarised in figures \ref{scatter}-\ref{s6}) have been established in terms of the cumulants, $S_n$, and so they hold even if there are other types of non-Gaussianity present. The only difference in this case is that it will be more complicated to translate the cumulants into $\fnl$-type parameters. For instance, $S_3$ will now comprise a mixture of local and non-local contributions
\be S_3 = \fnl\super{local} \mc{I}_1+\fnl\super{nonlocal}\mc{I}_2,\ee
where $\mc{I}_1$ and $\mc{I}_2$ are some integral expressions. See \cite{loverde,desjacques2} for the expressions for $\mc{I}_2$ in the case where non-Gaussianity is of the so-called folded or equilateral-triangle type.

\bbb

% CLARIFY the difference between cumulants and moments

Having understood how to produce well-defined non-Gaussian pdfs using the Edgeworth expansion, we shall now look at two applications, namely, the non-Gaussian prediction for abundances of clusters and voids. In what follows, we shall focus on the case where $\fnl=0$ and $\gnl>0$.

\section{Abundance of massive clusters}

Large-scale structures are sensitive to primordial non-Gaussianity on scales much smaller than the CMB (see \cite{desjacques2} for a recent review).
 On these scales, non-Gaussianity can manifest in the changes in cluster number count and its redshift dependence \citep{lucchin,robinson,loverde,oguri} as well as a scale-dependent halo bias \citep{dalal,matarrese,wands}. In this work, we use the Edgeworth approach, in its correct formalism, together with Press-Schechter theory to study the effect of non-zero $\gnl$ on the number density of massive clusters. Redshift dependence and the effects on the correlation function will be examined in a later publication.

\subsection{Press-Schechter theory}

Let $n(M)$ be the number density of collapsed objects of mass above $M$. Press-Schechter theory \citep{ps} gives the differential number density of collapsed objects as
\be  \diff{n}{M} = -2{\rho_m\over M}\diff{}{M}\int_{\delta_c /\sigma(M)}^{\infty}p(\nu,M)d\nu,\lab{massfun}\ee
where  $p(\nu,M)$ is the pdf smoothed by a window function containing mass $M$ and $\delta_c\approx1.686$ is the threshold overdensity for spherical collapse. %Note that the redshift dependence occurs in $\delta_c$ only.
For a non-Gaussian pdf\footnote{We note that there are a number of other formalisms for calculating the non-Gaussian contributions to the mass function. See \eg \cite{damico,maggiore1,maggiore2,maggiore3}.}
, \cite{grossi} suggest that a good fit to N-body simulations can be obtained by using the Press-Schechter mass function modified by the replacement
\be \delta_c\rightarrow 0.866\ff\delta_c,\lab{replace}\ee
(see \cite{maggiore3} for a possible theoretical origin.) We make this replacement in our calculations. 

\begin{figure}
\centering
\epsfig{file=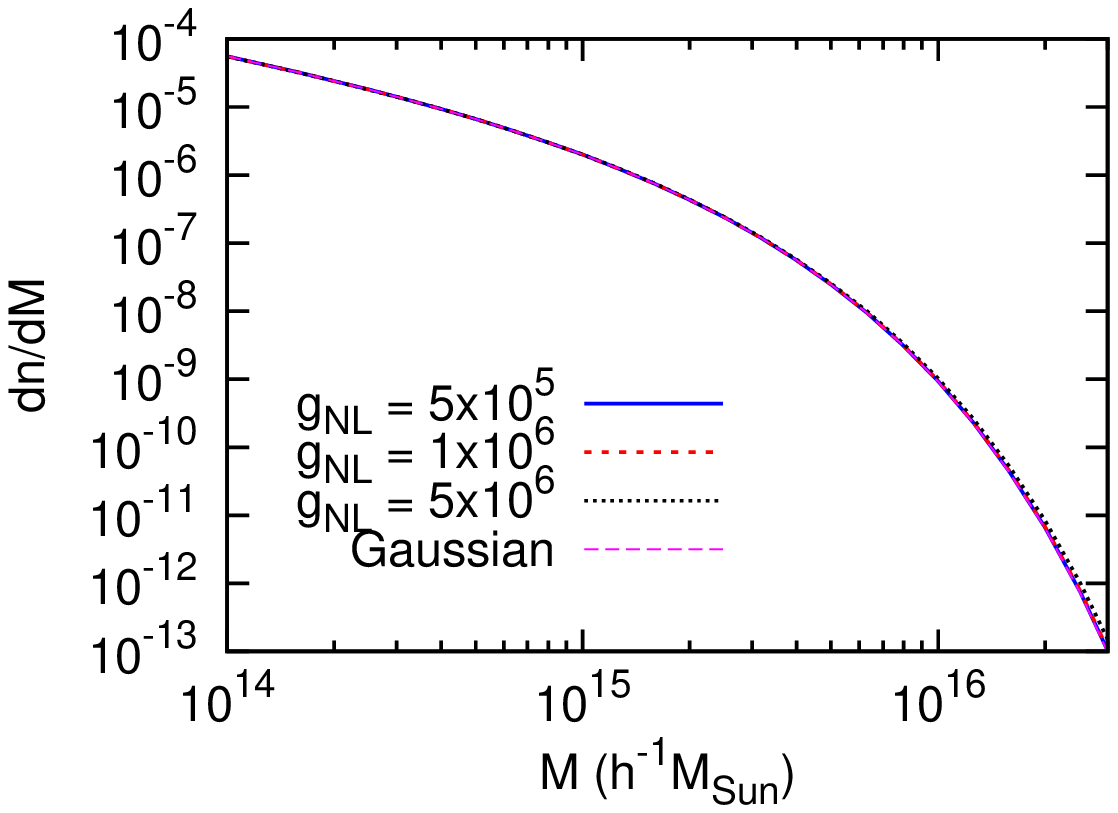, height= 8.6cm, angle = -90}\epsfig{file=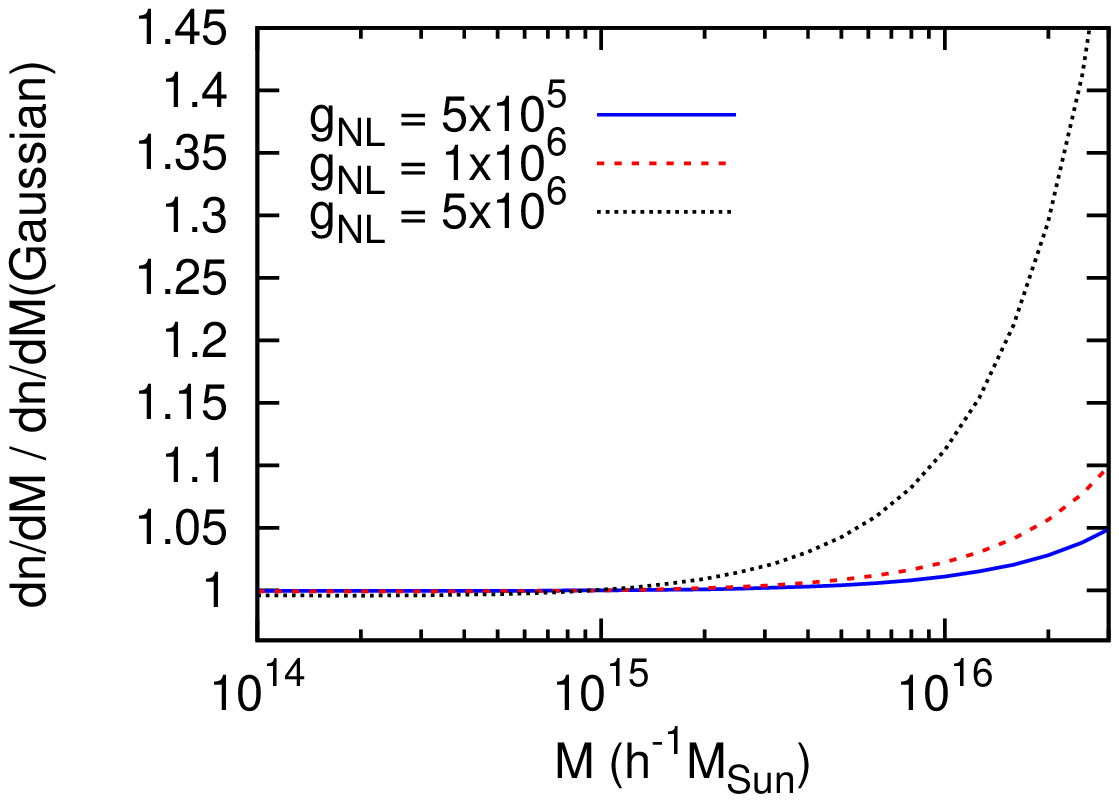, height= 8.6cm, angle = -90}
\caption{ \ii{Left}: The differential number density $dn/dM$ as a function of mass scale $M$ for models with $\gnl=5\times10^{5},1\times10^{6}$ and $5\times10^{6}$ ($\fnl=0$). The number density of massive clusters increases with $\gnl$. \ii{Right}: Ratios between the non-Gaussian and Gaussian number densities.}
\label{figcompare}
\end{figure}

Figure \ref{figcompare} shows the changes in $dn/dM$ for a range of non-Gaussian models with $\gnl=5\times10^{5},1\times10^{6}$ and $5\times10^6$ ($\fnl=0$ in all cases). In these calculations, we keep the Edgeworth expansion up to $10$ terms and check that $p(\nu)>0$ at least in the range $\nu\in[-20,20]$. Outside this range, the $p(\nu)$ is sufficiently small and the contribution to the cluster abundance on this mass scale is negligible (note that for the normal distribution, $N(20)\sim10^{-88}$). The values of $\gnl$ have been chosen to stay within the region of validity (see figure \ref{scatterlots}). In our case, we require $0\leq\sigma^2 S_4\lesssim0.6$, corresponding roughly to $0\lesssim\gnl\lesssim\mc{O}(10^8)$. 

The general effect of $\gnl>0$ is a boost in the number density of the most massive objects, although a significant boost requires the magnitude of $\gnl$ to exceed the CMB-derived bound of \cite{vielva}. For instance, abundance of objects of mass $\sim10^{16} M_\sun$ (corresponding to the most massive clusters) is increased by about 10\% for $\gnl=5\times10^{6}$. Nevertheless, values of $\gnl$ of this magnitude has recently been proposed by Enqvist \etal \citep{enqvist} to explain the observed excess of massive clusters. Until there is a larger  compilation of massive clusters, the possibility of non-Gaussianity with $\gnl\sim\mc{O}(10^6)$ remains a viable.

%%%%%%%%%%%%%%%%%%%%%%%%%%

%%%%%%%%%%%%%%%%%%%%%%%%%%

\subsection{Sensitivity to truncation}

% Graph showing pdf
Increasing the number of terms in the Edgeworth expansion does not change the pdf drastically. However, because the mass function is extremely sensitive to the exponential tail of the distribution, it is imperative that one keeps as many terms as practically possible in the Edgeworth expansion. Exactly how many terms are required will depend on a combination of factors such as the range of scales of interest or the redshift at which the calculations are made. 

\begin{figure}
\centering
\epsfig{file=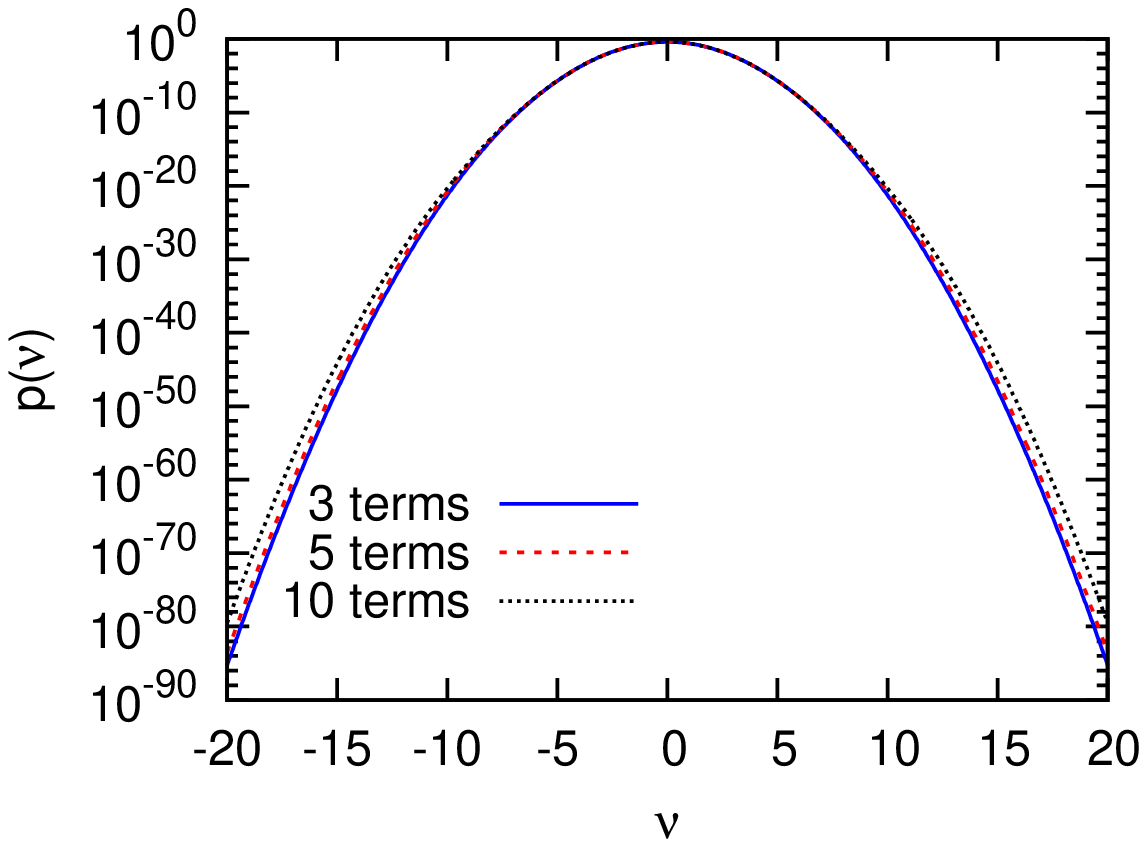, height= 8.5cm, angle = -90}
\epsfig{file=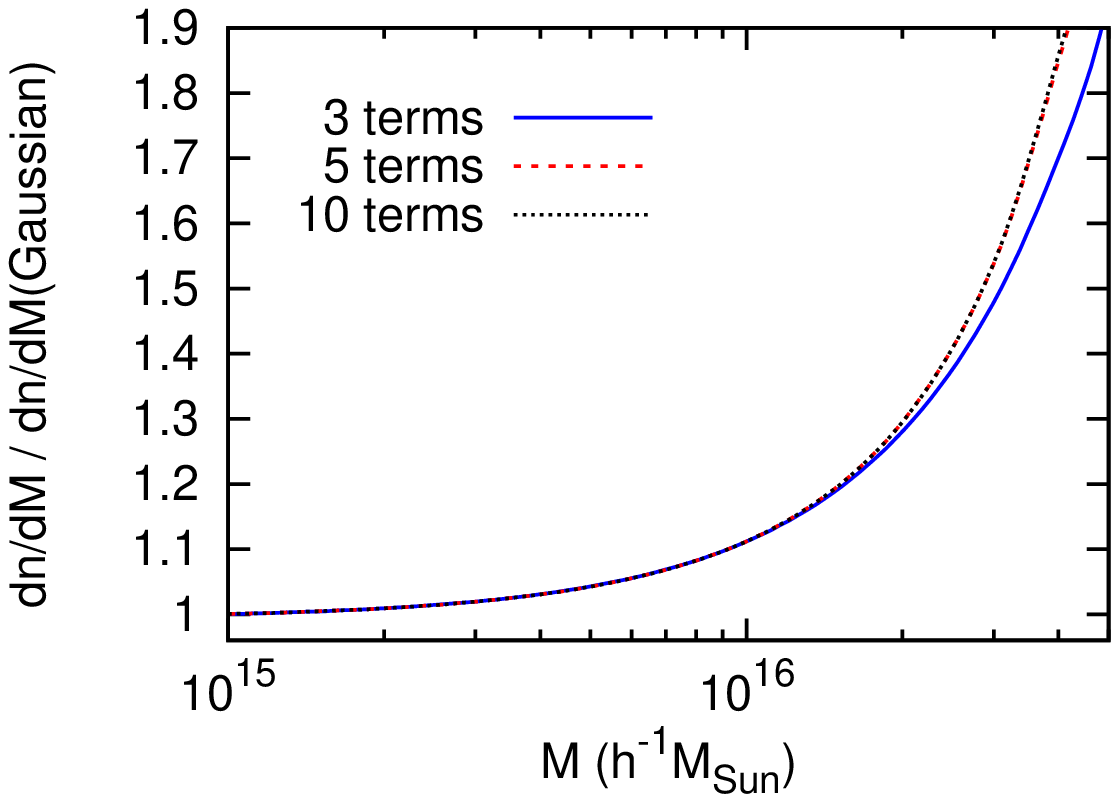, height= 8.9cm, angle = -90}
\caption{The sensitivity of the Edgeworth series to the truncation order \ii{Left:} the pdf $p(\nu)$ for models with $\gnl=5\times10^{6}$. Increasing the number of terms  in the Edgeworth expansion from 3 to 10 changes the pdf at high values of $|\nu|$ only by a tiny amount (note the logarithmic scale on the vertical axis). \ii{Right:} the effect of increasing the number of terms in the Edgeworth series on the cluster abundance $dn/dM$. A fifth (and higher) order expansion can give $dn/dM$ (shown here as a ratio to the Gaussian value) higher than a third-order truncation by up to $10\%$ at high $M$. Further increment in the number of terms has no discernible effect on the abundance in this mass range.}
\label{danger}
\end{figure}

% repeat for gnl only
Figure \ref{danger} demonstrates the sensitivity of the Edgeworth expansion to the truncation order. The panel on the left shows the pdf for a distribution with $\gnl=5\times10^{6}$ (with higher-order cumulants again equal $0$) and $R=8h^{-1}$Mpc. The various lines correspond to the number of terms in the Edgeworth expansion. For $\nu\sim\mc{O}(1)$, the pdfs lie almost exactly on top of one another, diverging only at the tail ends. However, when dealing with extreme-mass objects, keeping just a few terms in the Edgeworth series is inadequate, as seen in the panel on the right. Here, the ratio of the non-Gaussian number density $dn/dM$ and the Gaussian value can change by $10\%$ as the number of terms increases from 3 to 5 on mass scales beyond  $10^{16}M_\sun$. Comparing with figure \ref{figcompare}, we conclude that fitting the observed abundance of massive clusters using a low truncation order would lead to a spuriously high value of $\gnl$ and vice versa. 

We note that is likely that the sensitivity to the truncation order could increase significantly with redshift. We shall address this issue in a forthcoming work.

%%%%%%%%%%%%%%%%%%%%%%%%%%
%%%%%%%%%%%%%%%%%%%%%%%%%%

%%%%%%%%%%%%%%%%%%%%%%%%%%

%%%%%%%%%%%%%%%%%%%%%%%%%%
%%%%%%%%%%%%%%%%%%%%%%%%%%
%%%%%%%%%%%%%%%%%%%%%%%%%%

\section{Abundance of voids}

Primordial non-Gaussianity also changes the abundance of underdense regions, \iee cosmic voids. An estimate of void abundance can be computed by a simple extension of the Press-Schechter formalism \citep{kamionkowski,biswas}, although there are more sophisticated methods based on the void probability function \citep{white} or the eigenvalues of the tidal tensor \citep{doroshkevich,lam}. Presently, we shall use the Press-Schechter approach with the Edgeworth expansion to calculate the effect of $\gnl$ on the void abundance.

A void can be defined as an isolated region in which $-1\leq\delta\leq\delta_v$, where $\delta_v$ is some threshold underdensity. Simulations carried out by \cite{shandarin},\cite{park} and \cite{colberg} suggest $\delta_v\approx-0.8$. Linearly extrapolating this value to $z=0$ using the fitting formula of \cite{mo} gives $\delta_v=-2.75$.

Let $\mbox{Prob}_{<\delta_v}(\mb{x})$ be the probability that $\delta<\delta_v$ at $\mb x$. Since $\mbox{Prob}_{<\delta_v}=1-\mbox{Prob}_{>\delta_v}$, differentiating this expression with respect to $R$ shows that the analog of equation \ref{massfun} for voids can simply be obtained by the replacement $\delta_c\rightarrow\delta_v$ and a change in the overall sign.

%%%%%%%%%%%%%%%%%%%%%%%%%%

\begin{figure}
\centering
\epsfig{file=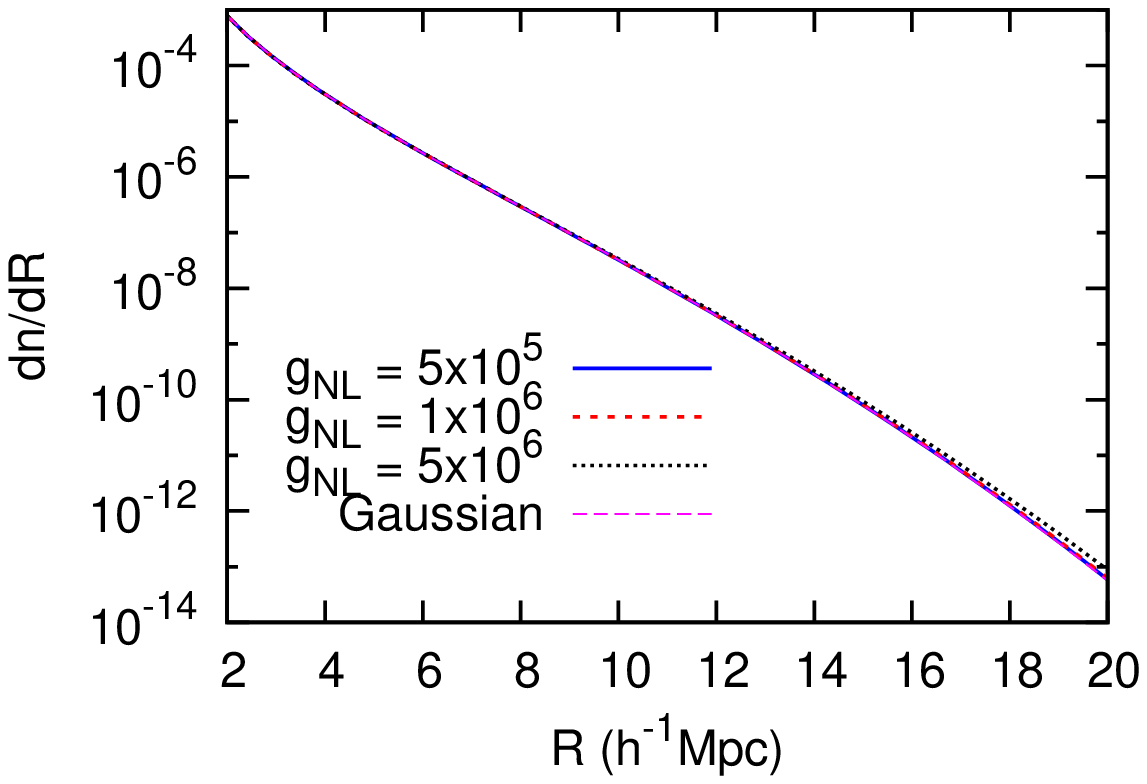, height= 7.7cm, angle = -90}
\epsfig{file=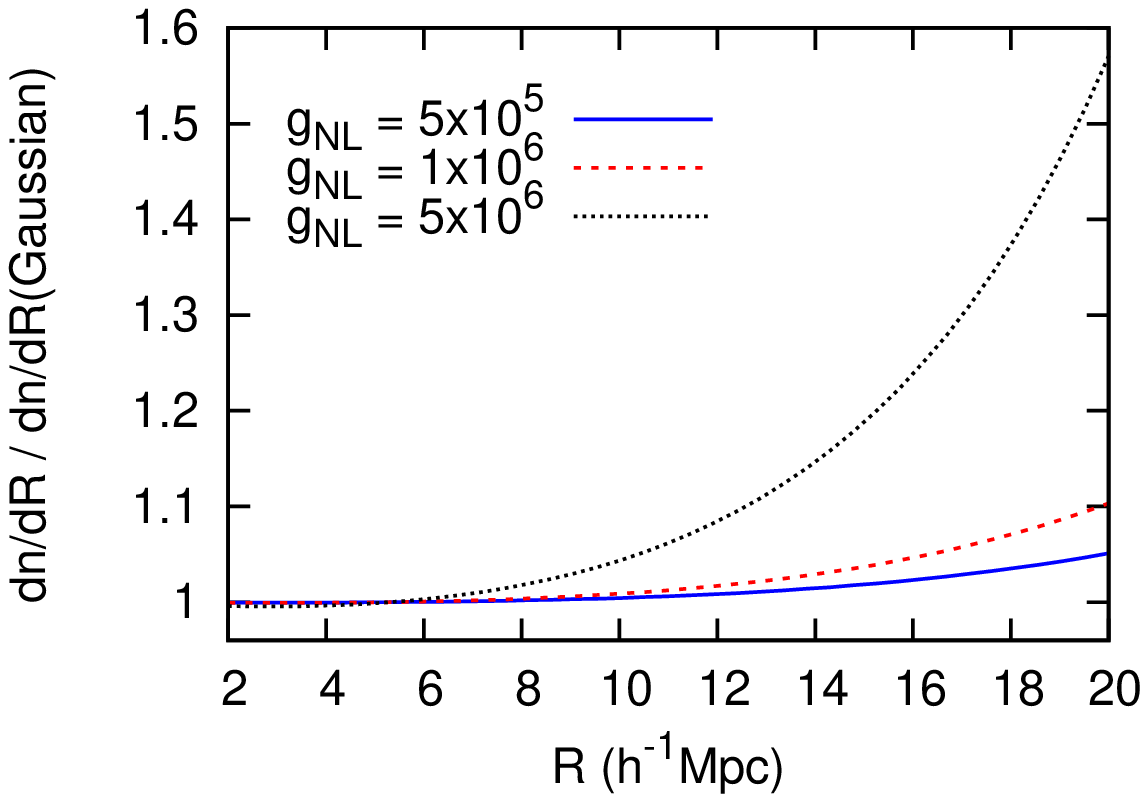, height= 7.8cm, angle = -90}
\caption{\ii{Left}: The differential void abundance $dn/dR$ as a function of scale $R$ for models with $\gnl=5\times10^{5},1\times10^{6}$ and $5\times10^{6}$ ($\fnl=0$). The number density of large voids increases with $\gnl$. \ii{Right}: Ratios between the non-Gaussian and Gaussian void abundances, showing the factor of enhancement. Voids appear to respond much more sensitively to $\gnl$ compared to clusters (see figure \ref{figcompare}).}
\label{void}
\end{figure} 

%%%%%%%%%%%%%%%%%%%%%%%%%%
%%%%%%%%%%%%%%%%%%%%%%%%%%
%%%%%%%%%%%%%%%%%%%%%%%%%%

Figure \ref{void} shows the differential number density $dn/dR$ for models in which $\gnl=5\times10^{5},1\times10^{6}$ and $5\times10^{6}$ , plotted against the smoothing scale $R$. By increasing $\gnl$, the void abundance is enhanced and responds much more sensitively than the cluster abundance (compare figures \ref{figcompare} and \ref{void}). For example, in the extreme case where $\gnl=5\times10^{6}$, at $R=20\ff h^{-1}\mbox{Mpc}$ ($M\approx10^{16}\ff h^{-1}M_\sun$), the enhancement in the differential abundance compared to the Gaussian prediction is roughly $10\%$ for clusters, but as large as $60\%$ for voids. This suggests that large voids may be a more sensitive probe of primordial non-Gaussianity than massive clusters, although a more careful calculation is needed to confirm this.

Another interesting observation is that $\gnl>0$ enhances both cluster and void abundances. This is in contrast with the effect of $\fnl>0$, which enhances the number density of clusters, but suppresses the number density of voids \citep{lam,kamionkowski}. Comparing the abundances of clusters and voids may offer a way to probe any asymmetry in the distribution of density fluctuations.

%%%%%%%%%%%%%%%%%%%%%%%%%%
%%%%%%%%%%%%%%%%%%%%%%%%%%
\section{Conclusions}

The key results in this paper are as follows

\begin{itemize}
\item We clarified the statistical meaning of the local non-Gaussianity parameters $\fnl$ and $\gnl$, which, at leading order, are proportional to the skewness and excess kurtosis of the distribution of density fluctuations. These relations are in the form of multi-dimensional integrals, which can be fitted by simple formulae \re{scale1} and \re{scale2}. They are accurate on the mass scale $10^{13}-10^{16}\ff M_\sun$.

\item We showed that the information in $\fnl$ and $\gnl$ is insufficient for a reconstruction of the pdf of density fluctuations. Using a theorem from the classical Hamburger moment problem, we showed that there is no positive pdf which deviates from Gaussianity only in the 3rd and 4th moments.

\item We studied the truncated Edgeworth series, emphasising that in this representation, $\fnl$ and $\gnl$ may not accurately reflect the skewness and excess kurtosis of the reconstructed pdf, especially for shorter truncations. We surveyed the skewness-kurtosis plane for regions of validity (\iee where the pdf is non-negative) for various truncations of the Edgeworth series. We proved that the Edgeworth expansion can represent a non-negative pdf if it is truncated at even order in $\sigma$, with the highest-order cumulant restricted to non-negative values. In terms of local non-Gaussianity, this means that the Edgeworth series cannot be used to represent models with nonzero $\fnl$ without considering nonzero $\gnl$ also. It also means that models with $\gnl<0$ are not representable by a truncated Edgeworth series unless the non-Gaussian deviation in the 6th moment (or higher) is known. 

\item Working with a 10th-order Edgeworth series, we calculated the effects of $\gnl$ on the cluster number density, $dn/dM$, using the Press-Schechter formalism (see figure \ref{figcompare}). The differential abundance of the most massive clusters can increase significantly if $\gnl\sim10^{6}$. We cautioned that the deduced value of $\gnl$ from large-scale structures may be spuriously high if the series is prematurely truncated.

\item Finally, we extended the Press-Schechter approach to compute the effects of $\gnl$ on the abundance of large voids. The void number density is enhanced much more sensitively compared to clusters (figure \ref{void}). This could be confirmed by a more sophisticated calculation (\eg using correlation functions to calculate the void probability distribution). We shall address this issue in a future publication.

\end{itemize}

%%%%%%%%%%%%%%%%%%%%%%%%%%%%%%%%%%%%%%%%%%%%%%%%%%%

\bbb

\centerline{\large{\sc{acknowledgments}}}

\bbb

SC is supported by Lincoln College, Oxford. We are grateful to Profs. N. Sugiyama and S. Zaroubi for early discussions that helped inspired this project, to Sergei Blinnikov for sharing his Petrov-Edgeworth code, to Lindsay King and Pedro Ferreira for reading the manuscript, and to Bob Scherrer and the referee for many helpful suggestions. Many thanks also to Shaun Hotchkiss, Sarah Shandera, Qing-Guo Huang, Laura Cay\'{o}n and Elizabeth Eardley for discussions that led to the improvement of the original version.

\bbb

\liner

\appendix

\bibliography{ng}

\begin{thebibliography}{56}
\expandafter\ifx\csname natexlab\endcsname\relax\def\natexlab#1{#1}\fi

\bibitem[{Akheizer(1965)}]{akheizer}
Akheizer, N.~I. 1965, The classical moment problem and related questions in
  analysis (New York: Hafner Publishing Company)

\bibitem[{Alishahiha {et~al.}(2004)Alishahiha, Silverstein, \&
  Tong}]{alishahiha}
Alishahiha, M., Silverstein, E., \& Tong, D. 2004, Phys. Rev. D, 70, 123505

\bibitem[{{Amendola}(2002)}]{amendola}
{Amendola}, L. 2002, \apj, 569, 595, arXiv:astro-ph/0107527

\bibitem[{Arkani-Hamed {et~al.}(2004)Arkani-Hamed, Creminelli, Mukohyama, \&
  Zaldarriaga}]{arkani-hamed}
Arkani-Hamed, N., Creminelli, P., Mukohyama, S., \& Zaldarriaga, M. 2004, JCAP,
  0404, 001, hep-th/0312100

\bibitem[{Balitskaya \& Zolotuhina(1988)}]{balitskaya}
Balitskaya, E.~O., \& Zolotuhina, L.~A. 1988, Biometrika, 75, 185

\bibitem[{Bartolo {et~al.}(2004{\natexlab{a}})Bartolo, Komatsu, Matarrese, \&
  Riotto}]{bartolo}
Bartolo, N., Komatsu, E., Matarrese, S., \& Riotto, A. 2004{\natexlab{a}},
  Phys. Rept., 402, 103

\bibitem[{Bartolo {et~al.}(2004{\natexlab{b}})Bartolo, Matarrese, \&
  Riotto}]{bartolo2}
Bartolo, N., Matarrese, S., \& Riotto, A. 2004{\natexlab{b}}, Phys. Rev. D, 69,
  043503

\bibitem[{{Bernardeau} \& {Kofman}(1995)}]{bernardeau}
{Bernardeau}, F., \& {Kofman}, L. 1995, \apj, 443, 479, arXiv:astro-ph/9403028

\bibitem[{Biswas {et~al.}(2010)Biswas, Alizadeh, \& Wandelt}]{biswas}
Biswas, 1, R., Alizadeh, E., \& Wandelt, B.~D. 2010, 1002.0014

\bibitem[{{Blinnikov} \& {Moessner}(1998)}]{blinnikov}
{Blinnikov}, S., \& {Moessner}, R. 1998, \aaps, 130, 193,
  arXiv:astro-ph/9711239

\bibitem[{{Byrnes} \& {Choi}(2010)}]{byrnes}
{Byrnes}, C.~T., \& {Choi}, K. 2010, 1002.3110

\bibitem[{Cappi \& Maurogordato(1995)}]{cappi}
Cappi, A., \& Maurogordato, S. 1995, \apj, 438, 507

\bibitem[{Cayon {et~al.}(2010)Cayon, Gordon, \& Silk}]{cayon}
Cayon, L., Gordon, C., \& Silk, J. 2010, 1006.1950

\bibitem[{Chen(2005)}]{chen2}
Chen, X. 2005, Phys. Rev., D72, 123518

\bibitem[{Chen(2010)}]{chen}
------. 2010, 1002.1416

\bibitem[{Chen {et~al.}(2007)Chen, Huang, Kachru, \& Shiu}]{chen3}
Chen, X., Huang, M.-x., Kachru, S., \& Shiu, G. 2007, JCAP, 0701, 002,
  hep-th/0605045

\bibitem[{Colberg {et~al.}(2008)}]{colberg}
Colberg, J.~M., {et~al.} 2008, 0803.0918

\bibitem[{Dalal {et~al.}(2008)Dalal, Dore, Huterer, \& Shirokov}]{dalal}
Dalal, N., Dore, O., Huterer, D., \& Shirokov, A. 2008, Phys. Rev., D77,
  123514, 0710.4560

\bibitem[{{D'Amico} {et~al.}(2010){D'Amico}, {Musso}, {Nore{\~n}a}, \&
  {Paranjape}}]{damico}
{D'Amico}, G., {Musso}, M., {Nore{\~n}a}, J., \& {Paranjape}, A. 2010,
  1005.1203

\bibitem[{Desjacques \& Seljak(2010)}]{desjacques2}
Desjacques, V., \& Seljak, U. 2010, 1006.4763

\bibitem[{Desjacques {et~al.}(2008)Desjacques, Seljak, \& Iliev}]{desjacques}
Desjacques, V., Seljak, U., \& Iliev, I. 2008, 0811.2748

\bibitem[{{Doroshkevich}(1970)}]{doroshkevich}
{Doroshkevich}, A.~G. 1970, Astrofizika, 6, 581

\bibitem[{Draper \& Tierney(1972)}]{draper}
Draper, N.~R., \& Tierney, D.~E. 1972, Biometrika, 59, 463

\bibitem[{Eisenstein \& Hu(1998)}]{eisenstein}
Eisenstein, D.~J., \& Hu, W. 1998, \apj, 496, 605

\bibitem[{{Enqvist} {et~al.}(2010){Enqvist}, {Hotchkiss}, \&
  {Taanila}}]{enqvist}
{Enqvist}, K., {Hotchkiss}, S., \& {Taanila}, O. 2010, 1012.2732

\bibitem[{Grossi {et~al.}(2009)}]{grossi}
Grossi, M., {et~al.} 2009, Mon. Not. Roy. Astron. Soc., 398, 321

\bibitem[{Jondeau \& Rockinger(2001)}]{jondeau}
Jondeau, E., \& Rockinger, M. 2001, Journal of Economic Dynamics and Control,
  25, 1457

\bibitem[{{Juszkiewicz} {et~al.}(1995){Juszkiewicz}, {Weinberg},
  {Amsterdamski}, {Chodorowski}, \& {Bouchet}}]{juszkiewicz}
{Juszkiewicz}, R., {Weinberg}, D.~H., {Amsterdamski}, P., {Chodorowski}, M., \&
  {Bouchet}, F. 1995, \apj, 442, 39, arXiv:astro-ph/9308012

\bibitem[{Kamionkowski {et~al.}(2009)Kamionkowski, Verde, \&
  Jimenez}]{kamionkowski}
Kamionkowski, M., Verde, L., \& Jimenez, R. 2009, JCAP, 0901, 010

\bibitem[{Kjeldsen(1993)}]{kjeldsen}
Kjeldsen, T.~H. 1993, Historia Mathematica, 20, 19

\bibitem[{Komatsu {et~al.}(2010)}]{komatsu}
Komatsu, E., {et~al.} 2010, 1001.4538

\bibitem[{{Kurokawa} {et~al.}(1999){Kurokawa}, {Morikawa}, \&
  {Mouri}}]{kurokawa}
{Kurokawa}, T., {Morikawa}, M., \& {Mouri}, H. 1999, \aap, 344, 1

\bibitem[{Lahav \& Liddle(2010)}]{lahav+}
Lahav, O., \& Liddle, A.~R. 2010, 1002.3488

\bibitem[{Lam {et~al.}(2009)Lam, Sheth, \& Desjacques}]{lam}
Lam, T.~Y., Sheth, R.~K., \& Desjacques, V. 2009, 0905.1706

\bibitem[{Langlois {et~al.}(2008)Langlois, Renaux-Petel, Steer, \&
  Tanaka}]{langlois}
Langlois, D., Renaux-Petel, S., Steer, D.~A., \& Tanaka, T. 2008, Phys. Rev.
  Lett., 101, 061301

\bibitem[{LoVerde {et~al.}(2008)LoVerde, Miller, Shandera, \& Verde}]{loverde}
LoVerde, M., Miller, A., Shandera, S., \& Verde, L. 2008, JCAP, 0804, 014

\bibitem[{{Lucchin} \& {Matarrese}(1988)}]{lucchin}
{Lucchin}, F., \& {Matarrese}, S. 1988, \apj, 330, 535

\bibitem[{Maggiore \& Riotto(2009{\natexlab{a}})}]{maggiore1}
Maggiore, M., \& Riotto, A. 2009{\natexlab{a}}, 0903.1249

\bibitem[{Maggiore \& Riotto(2009{\natexlab{b}})}]{maggiore3}
------. 2009{\natexlab{b}}, 0903.1251

\bibitem[{Maggiore \& Riotto(2010)}]{maggiore2}
------. 2010, \apj, 717, 515, 0903.1250

\bibitem[{Matarrese \& Verde(2008)}]{matarrese}
Matarrese, S., \& Verde, L. 2008, \apj, 677, L77

\bibitem[{Mo \& White(1996)}]{mo}
Mo, H.~J., \& White, S. D.~M. 1996, Mon. Not. Roy. Astron. Soc., 282, 347

\bibitem[{Oguri(2009)}]{oguri}
Oguri, M. 2009, Phys. Rev. Lett., 102, 211301

\bibitem[{Park \& Lee(2007)}]{park}
Park, D., \& Lee, J. 2007, Phys. Rev. Lett., 98, 081301

\bibitem[{Petrov(1975)}]{petrov}
Petrov, V. 1975, Ergebnisse der Mathematik und ihrer Grenzgebiete, Vol.~82,
  Sums of independent random variables (Berlin: Springer-Verlag)

\bibitem[{Press \& Schechter(1974)}]{ps}
Press, W.~H., \& Schechter, P. 1974, \apj, 187, 425

\bibitem[{Rigopoulos {et~al.}(2006)Rigopoulos, Shellard, \& van
  Tent}]{rigopoulos}
Rigopoulos, G.~I., Shellard, E. P.~S., \& van Tent, B. J.~W. 2006, Phys. Rev.,
  D73, 083522

\bibitem[{{Robinson} \& {Baker}(2000)}]{robinson}
{Robinson}, J., \& {Baker}, J.~E. 2000, \mnras, 311, 781,
  arXiv:astro-ph/9905098

\bibitem[{Sasaki {et~al.}(2006)Sasaki, V\"aliviita, \& Wands}]{sasaki}
Sasaki, M., V\"aliviita, J., \& Wands, D. 2006, Phys. Rev. D, 74, 103003

\bibitem[{Sefusatti {et~al.}(2009)Sefusatti, Liguori, Yadav, Jackson, \&
  Pajer}]{sefusatti}
Sefusatti, E., Liguori, M., Yadav, A. P.~S., Jackson, M.~G., \& Pajer, E. 2009,
  JCAP, 0912, 022

\bibitem[{Shandarin {et~al.}(2006)Shandarin, Feldman, Heitmann, \&
  Habib}]{shandarin}
Shandarin, S., Feldman, H.~A., Heitmann, K., \& Habib, S. 2006, Mon. Not. Roy.
  Astron. Soc., 367, 1629

\bibitem[{Shohat \& Tamarkin(1963)}]{shohat}
Shohat, J., \& Tamarkin, J. 1963, Mathematical Surveys and Monographs, Vol.~1,
  The Problem of Moments (American Mathematical Society)

\bibitem[{{Vielva} \& {Sanz}(2010)}]{vielva}
{Vielva}, P., \& {Sanz}, J.~L. 2010, \mnras, 404, 895

\bibitem[{Wands \& Slosar(2009)}]{wands}
Wands, D., \& Slosar, A. 2009, Phys. Rev., D79, 123507

\bibitem[{Weinberg(2008)}]{weinberg}
Weinberg, S. 2008, Cosmology (Oxford University Press)

\bibitem[{{White}(1979)}]{white}
{White}, S.~D.~M. 1979, \mnras, 186, 145

\end{thebibliography}

\end{document}